\documentclass[preprint,showpacs,preprintnumbers,nofootinbib,amsmath,amssymb]
{revtex4-1}
\usepackage{graphics,epsfig,subfigure}
\usepackage{epstopdf}
\usepackage{color}
\usepackage{graphicx}

\newcommand{\mysection}[1]{\section{#1}
\renewcommand{\theequation}{\thesection.\arabic{equation}}
\setcounter{equation}{0}}

\begin{document}
\renewcommand{\baselinestretch}{1.3}
\newcommand\be{\begin{equation}}
\newcommand\ee{\end{equation}}
\newcommand\ba{\begin{eqnarray}}
\newcommand\ea{\end{eqnarray}}
\newcommand\nn{\nonumber}
\newcommand\fc{\frac}
\newcommand\lt{\left}
\newcommand\rt{\right}
\newcommand\pt{\partial}
\newcommand\tc{\textcolor[rgb]{1,0,0}}

\newcommand{\rL}{\rho_\Lambda}
\newcommand{\pL}{p_\Lambda}
\newcommand{\drL}{\dot{\rho}_\Lambda}
\newcommand{\dprL}{\rho'_\Lambda}
\newcommand{\CC}{\Lambda}
\newcommand{\newtext}[1]{{\textcolor{red}{#1}}}
\newcommand{\rLo}{\rho_{\CC 0}}

\newcommand{\rmm}{\rho_{m}}
\newcommand{\rcdm}{\rho_{cdm}}
\newcommand{\rb}{\rho_{b}}
\newcommand{\drm}{\rho'_{m}}
\newcommand{\rmo}{\rho_{m 0}}
\newcommand{\rr}{\rho_{r}}
\newcommand{\rro}{\rho_{r 0}}
\newcommand{\rco}{\rho_{c0}}

\newcommand{\Omo}{\Omega_{m 0}}
\newcommand{\OLo}{\Omega_{\CC 0}}
\newcommand{\Oro}{\Omega_{r 0}}

\newcommand{\V}{\cal V}
\newcommand{\A}{\cal A}

\newcommand{\SV}{S_{\cal V}}
\newcommand{\SA}{S_{\cal A}}

\newcommand{\SVp}{S'_{\cal V}}
\newcommand{\SAp}{S'_{\cal A}}

\newcommand{\SVpp}{S''_{\cal V}}
\newcommand{\SApp}{S''_{\cal A}}

\newcommand{\ha}{\hat{a}}
\newcommand{\astar}{a_{*}}
\newcommand{\tHI}{\tilde{H}_I}
\newcommand{\trI}{\tilde{\rho}_I}
\newcommand{\tTI}{\tilde{T}_I}

\title{Particle and entropy production in the Running Vacuum Universe}
\author{Joan Sol\`a Peracaula$^{1}~$\footnote{sola@fqa.ub.edu} and  Hao Yu$^{1,2}~$\footnote{yuhao@fqa.ub.edu}\\
$^1$Departament de F\'\i sica Qu\`antica i Astrof\'\i sica,\\
and Institute of Cosmos Sciences (ICCUB)\\
Universitat de Barcelona, Avinguda Diagonal 647 E-08028 Barcelona, Catalonia, Spain\\
$^2$Institute of Theoretical Physics, Lanzhou University, Lanzhou 730000, China
\vspace{1cm}}

\begin{abstract}
We study particle production and the corresponding entropy increase in the context of cosmology with dynamical vacuum. We focus on the particular form that has been called ``running vacuum model'' (RVM), which is known to furnish a successful description of the overall current observations at a competitive level with the concordance $\Lambda$CDM model. It also provides an elegant global explanation of the cosmic history from a non-singular initial state in the very early universe up to our days and further into the final de Sitter era. The model has no horizon problem and offers an alternative explanation for the early inflation and its graceful exit, as well as a powerful mechanism for generating the large entropy of the current universe.  The energy-momentum tensor of  matter is generally non-conserved in such context owing to particle creation or annihilation. We analyze general thermodynamical aspects of particle and entropy production in the RVM. We first study the entropy of particles in the comoving volume during the early universe and late universe. Then, in order to obtain a more physical interpretation, we pay attention to the entropy contribution from the cosmological apparent horizon, its interior and its surface. On combining the inner volume entropy with the entropy on the horizon, we elucidate with detailed calculations whether the evolution of the entropy of the RVM universe satisfies the Generalized Second Law of Thermodynamics. We find it is so and we prove that the essential reason for it is the existence of a positive cosmological constant.
\end{abstract}


\maketitle
\tableofcontents
\newpage
\mysection{Introduction}
The source of particles in the universe and the transformations between them have always been intriguing questions. We know that the origin of particles could be closely related to the starting of our universe, but people have widely different views about what is the ultimate mechanism, see e.g.\,\cite{WDW,Ford:1978ip,Vilenkin,HartleHawking1983,Linde1984,Rubakov1984,Grib2000,Zecca2012}. In the context of quantum cosmology, for instance  (see \,\cite{Vilenkin} and references therein), the universe is described by a global wave function rather than the classical spacetime.  Such a wave function of the universe should satisfy the Wheeler-DeWitt equation~\cite{WDW}, in which the Hamiltonian acting on the wave function is equal to zero.  Quantum cosmology actually suggests that the universe can be created spontaneously out of nothing, that is to say, from a state without matter nor space or time. This would solve the
singularity problem in a natural way, but we still lack of a consistent theory of spontaneous creation ``ex nihilo''. Some other mainstream cosmological models such as inflation\,\cite{Starobinsky1980,Guth1981,Linde1982} have their own ideas for explaining particle production and baryon asymmetry e.g. through reheating after inflation\,\cite{Dolgov1982,Abbot1982}. However, the mechanisms of inflation and reheating are manifold~~\cite{KolbTurner:1990,Linde1990,reviewLinde2014,Rubakov:2018} and we still have a long way to go before we can satisfactorily explain the ultimate source of particles in the universe. The standard $\CC$CDM model~\cite{LCDM1,LCDM2}, or ``concordance'' model of cosmology, does not have itself an explanation.

The types and proportions of particles in the early universe are completely different from our current universe. The emergence of new particles involves the exchange of energy between them, but in the process it can also participate the vacuum and its decay into particles or the annihilation of particles into the vacuum. The energy exchange we refer here works through the production or annihilation of particles at a macroscopic scale (the universe) and will be treated mainly on thermodynamical grounds. Therefore, the entropy description will be central in our approach. We will not address the microscopic details about how particles collide, the conditions of the energy exchange, the collision cross sections and so on. This would imply an exceedingly model-dependent description with many parameters.

The earliest study of particle production in cosmology dates back to the 1960s. The pioneering works of Parker and collaborators on time varying gravitational backgrounds are very representative~\cite{Parker:1968mv,Parker:1969au,Parker:1972kp}. Because the energy of the field is not conserved, its action is explicitly time-dependent and its quantization amounts to particle production. This semiclassical approach is based on quantum field theory (QFT) in curved spacetime\,\cite{BirrellDavis,ParkerToms}.  Since then, the research on particle production developed rapidly, especially its applications in cosmology -- see ~\cite{Sexl:1969ix,ZeldovichStarobinsky1972,Tryon:1973xi,Grib:1976pw,Gribosky:1985dz,
Hu:1993gm,Abramo:1996ip,Steigman2009,Lima2010,Paliathanasis:2016dhu}, for example. Among these applications, the thermodynamics of particle production is perhaps the most widely studied problem~\cite{Ranft:1970xu,Gibbons:1977mu,Kodama:1981jw,Hu:1986jj,
Kandrup:1988sg,Calvao:1991wg,Zimdahl96,RoyMaartens1996}. In Refs.~\cite{Prigogine:1986zz,Prigogine:1988zz,Prigogine:1989zz}, the Second Law of Thermodynamics is used to constrain particle production in cosmology within specific contexts. Since the Second Law  is a concept related to entropy, the notion of specific entropy (the entropy associated to a single particle) acquires a special significance  when particle production is involved in the evolution of the universe~\cite{Calvao:1991wg}. The study of Friedman-Lema\^{\i}tre-Robertson-Walker (FLRW)-type cosmology with adiabatic matter creation was discussed in \cite{LimaTrodden1996,LimaGermanoAbramo1996}, including some thermodynamic aspects of it in \cite{Lima1996,Graef:2013iia} and  applications to inflation~\cite{Ford:1986sy,Traschen:1990sw,Abramo:1996ip,
Gunzig:1997tk,Peebles:1998qn,Zimdahl96}. In recent years, the thermodynamic concepts have also been applied to modified gravity, scalar-tensor and Horndeski theories \cite{Harko:2014, Harko:2015,Yu:2018qzl}.

Being the production or annihilation of particles an energy-conversion process it can be characterized by the energy-momentum tensor of the corresponding matter. For General Relativity (GR) and minimally coupled theories of gravity, in general, the energy-momentum tensor of matter satisfies $\nabla_\mu T^{\mu\nu}=0$, and therefore there is no particle production and annihilation at a macroscopic scale. However, for more general cases, in particular for interactive theories of gravity and matter, the equation $\nabla_\mu T^{\mu\nu}=0$  does no longer apply for the matter part, and one can assume that there exists production or annihilation of matter.  This situation occurs e.g. within the class of GR-like theories of gravity in which the vacuum energy density $\rL=\CC/(8\pi G)$ evolves with the cosmic expansion, where $\CC$ is the cosmological term and $G$ is Newton's gravitational coupling.  The idea of a time-evolving $\CC$  is old enough, see e.g. ~\cite{Ozer:1985ws,Bertolami86,Freese1987,Carvalho:1991ut,Lima:1994gi}, but its implementation in practice has been changing significantly over time. The old models\,\cite{Overduin1998} are essentially phenomenological of nature, with little or no connection whatsoever with any fundamental theory.

In a more theoretical vein, we have the attempts to connect the evolution of the vacuum energy density  more closely with QFT in curved spacetime\,\cite{ShapSol,Fossil07}, and in particular also with the effective action of  Supergravity inflationary models\,\cite{RVM-SUGRA}. Along these lines we have the idea of the  `running vacuum model' (RVM), see \cite{JSPRev2013,Sola:2015rra} and references therein. Recently, a possible connection of the latter with the effective action of the bosonic gravitational multiplet  of string theory has been put forward\,\,\cite{Anomaly2019a,GRF2019}. Within the RVM  one can construct unified models of the cosmological evolution, in which vacuum plays a dynamical role with matter. It can provide a global picture of the universe evolution from inflation to the present days ~\cite{Lima:2012mu,Perico:2013mna}, see also \,\cite{Sola:2015rra} for a comprehensive presentation.  In fact one can mimic the particle production in a varying gravitational field. In such context, the  study of entropy production can be of great interest since there is a continuous interplay between matter and vacuum ~\cite{MimosoPavon2013,Lima:2014hia,GRF2015,Lima:2015mca,EspinozaPavon2019}. Phenomenologically, the class of RVM models show very good consistency with the observational data, see Refs.~\cite{Gomez-Valent:2014rxa,Sola:2015wwa,Sola:2016jky,Sola:2017znb,
Sola:2016hnq,Sola:2016ecz,Sola:2017jbl,Rezaei2019,Geng:2017apd,Perico:2016kbu}.
Moreover, it has been shown that some tensions of the concordance  $\CC$CDM model with the data can be alleviated with the option of dynamical vacuum or mimicking it, see e.g.\,\cite{JSPTensions,PDU2019,BDLett2019}. Different sorts of dynamical vacuum models (DVMs) can be of interest, see e.g.  \cite{Li2016,Valentino2016,Valentino2017,Costa2017,Martinelli2019,ParkRatra2018,Yang2019} and references therein, including nonparametric approaches\,\cite{GBZHao2017}.  Here, however, we wish to emphasize mainly on the thermodynamical aspects of these models, such as particle production and entropy increase.

In this work, we discuss general aspects of particle and entropy production within the RVM by considering the entire evolution of the universe from the inflationary epoch to the future de Sitter era\,\cite{Lima:2012mu,Perico:2013mna,JSPRev2013,Sola:2015rra,GRF2015}. We mainly focus on the study of the particle's entropy evolution and check if it satisfies the thermodynamic requirements concerning the Second Law. We separately address the entropy evolution from two different perspectives: i) the entropy of the  comoving volume,  and ii)  the total entropy in the presence of the apparent horizon, obtained from the sum of the volume plus the area contributions. This allows us to test explicitly the accomplishment of the Generalized Second Law (GSL) of thermodynamic  for the running vacuum universe. This aspect of our study is particularly interesting since it explores the holographic implications of the entropy in the presence of dynamical vacuum. For related studies, see e.g.\,\cite{KomatsuKimura,Komatsu} and references therein.

The paper is organized as follows: Sec.~\ref{sec2} is devoted to review basic facts of the thermodynamic framework in cosmology. Dynamical vacuum models are introduced  in Sec.~\ref{secDVM}. In Sec.~\ref{RVM}, we present the cosmological solutions for the RVM in the early and late universe. In Sections~\ref{ProductionCurrentUniverse} and \ref{ProductionEarlyUniverse}, we study particle and entropy production  in the comoving volume approach during the early and late time universe, respectively.
In Sec.~\ref{generalizedthermodynamiclaw}, we  focus on horizons, entropy and the GSL,  and discuss whether the entropy evolution for the RVM is  in line with the GSL expectations.  Sec.\,\ref{sec:Discussion}, acting as a kind of epilogue, discusses the lengthy path of the universe towards safe thermodynamical equilibrium in the context of the GSL. The main conclusions of this work are rendered in Sec.~\ref{sec7}. {Two appendices are also included. In Appendix A we rederive a relation of the text.  Appendix B extends our study for a generalized version of the RVM  and shows that the main conclusions remain intact.}

\mysection{Thermodynamics in expanding universe: basic formalism}
\label{sec2}
In this section, we introduce some basic thermodynamic formulas that will be used in our discussion  in subsequent sections. Because we will consider the vacuum dynamics in a cosmological context it proves useful to review the thermodynamic formulae in a cosmological spacetime and see what are the implied modifications. In such geometric arena we are assumed to have all the relevant ingredients, some of them in interaction, such as baryons, cold dark matter (CDM), radiation (made out of photons and neutrinos) and other (non-material) forms of energy, such as vacuum energy or in general dark energy (DE). They all participate in the thermodynamical analysis.

\subsection{Particle production and entropy flow}

In view of the cosmological principle\,\cite{LCDM2}, such spacetime is characterized by global homogeneity and isotropy and hence by the FLRW metric. For the spatially  flat case (to which we shall restrict our considerations, unless stated otherwise), the FLRW line element of spacetime $ds^2=g_{\mu\nu}dx^\mu dx^\nu$ adopts a particularly simple form:
\begin{equation}\label{eq:FLRWmetric}
ds^2=-c^2dt^2+a^2(t)(dx^2+dy^2+dz^2)\,,
\end{equation}
where the scale factor $a=a(t)$ is evolving with the cosmic time (see the next section for more details). Hereafter we assume natural units $c=\hbar=1$.

Let the number density of particles in FLRW spacetime be labeled as $n$. Combined with the four velocity of a comoving observer $u^\alpha$, it can be used to describe the flow of particles: $n^\alpha=n u^\alpha$. When there exists particle production (or annihilation), we can define a variable $\psi$ to characterize the non-conservation of the particle flow, i.e. $\nabla_\alpha n^\alpha=\psi$ ($\psi>0$ corresponds to a source of particles whereas $\psi<0$ to a sink, where particles disappear). The particle production rate of a given species of particles is then defined as follows:
\begin{eqnarray}
\Gamma=\frac{\psi}{n}.\label{particlerate}
\end{eqnarray}
The entropy flow correlating with the particle flow can be expressed as
\begin{eqnarray}
s^\alpha=\sigma n^\alpha=\sigma  n u^\alpha =s   u^\alpha,\label{entropydefination}
\end{eqnarray}
where $\sigma$ is the specific particle entropy (the entropy of an individual particle). In the comoving frame, we have $u^\alpha=(1,{\bf 0})$ and hence for $\alpha=0$ we obtain from (\ref{entropydefination}) the entropy density in that frame: $s=n\sigma$. Phrased in terms of (\ref{entropydefination}), the Second Law of Thermodynamics tells us that if the system is an isolate system, the total entropy flow always increases until it reaches equilibrium, i.e. $\nabla_\alpha s^\alpha\geq0$.

The non-conservation of the particle flux, i.e. $ \nabla_\alpha n^\alpha=\psi$, can be written in a more explicit form as follows:
\begin{eqnarray}
\dot{n}+\theta n=\psi=n\Gamma\,,\label{balanceLaw}
\end{eqnarray}
where a dot denotes differentiation with respect to the cosmic time.
The above expression gives the evolution of the particle number density (in the presence of the particle production rate $\Gamma$). In it
the parameter $\theta=\nabla_\alpha u^\alpha$ provides the expansion scalar of the fluid,  i.e. the
rate of change of the comoving volume during the expansion. For FLRW spacetime (\ref{eq:FLRWmetric}), one easily finds
\begin{equation}\label{eq:expansionscalar}
  \theta=\nabla_\alpha u^\alpha=\frac{1}{\sqrt{-g}}\partial_\alpha\left(\sqrt{-g} u^\alpha\right)=\frac{1}{a^3}\frac{d a^3}{dt}=3 H\,,
\end{equation}
where $g=-a^6$ is the determinant of the FLRW metric and  $H=\dot{a}/a$  is the Hubble rate.  It is interesting to note  that Eq.\,(\ref{balanceLaw}) allows us to obtain an alternative definition for the production rate in the number of created particles. Indeed, the total number of particles per comoving volume $a^3$ is $N=n a^3$, which satisfies
\begin{eqnarray}
\frac{\dot N}{N}=\frac{\dot n a^3+3n\dot a a^2}{na^3}
=\frac{\dot n +3Hn}{n}\,.\label{rateofchange}
\end{eqnarray}
Using now the balance law (\ref{balanceLaw}), this equation can be written in the compact form
\begin{eqnarray}
\Gamma=\frac{\dot N}{N}\,.\label{GammaAlternative}
\end{eqnarray}
As we can see, this provides an alternative definition of particle production rate $\Gamma$ (as the relative variation of the total number of particles with the cosmic time), which is more in accordance with intuition than the original definition in Eq.~(\ref{particlerate}).

The balance law (\ref{balanceLaw}) also allows us to compute the divergence of the entropy flow (\ref{entropydefination}):
\begin{eqnarray}
 \nabla_\alpha s^\alpha=n \dot{\sigma} +n \sigma \Gamma\,.\label{divergenceSflow}
\end{eqnarray}
As expected, if the specific particle entropy remains constant ($\dot{\sigma} =0$) and there is no particle production ($\Gamma=0$), then $ \nabla_\alpha s^\alpha=0$ and we are in thermodynamical equilibrium. If any or both of these conditions are not satisfied, the Second Law of Thermodynamics requires that the balance between the two terms yields $\nabla_\alpha s^\alpha\geq0$ until equilibrium is eventually reached. Using the last relation in \eqref{entropydefination}, we can write the alternative form
\begin{equation}\label{eq:nablas2}
  \nabla_\alpha s^\alpha=\nabla_\alpha \left(s   u^\alpha\right)=\dot{s}+3H s=\frac{1}{a^3}\frac{d(s a^3)}{dt}=\frac{1}{a^3}\frac{dS}{dt}\,,
\end{equation}
which does not involve the parameters $\sigma$ and $\Gamma$. The previous relation expresses the entropy flow just in terms of the time variation of the total entropy  $S=s a^3$ in a comoving volume.

We can further characterize the entropy change with the help of the thermodynamic potentials. Although thermodynamics was not originally formulated in the cosmological context, it is usually assumed that it can be applied to any physical volume element $V$ in the expanding universe~\cite{KolbTurner:1990,Rubakov:2018}. Thus, the First Law of thermodynamics for such volume can be expressed in the Gibbs form as follows:
\begin{equation}\label{eq:FirstLawThermo}
TdS=dU+pdV-\sum_i\mu_i dN_i\,,
\end{equation}
in which $T$ is the temperature,  $S$ is the total entropy,  $U$ is the total internal energy and $\mu_i\ (i=1,2,...)$ are the chemical potentials for the different components stored in the volume $V$. {Finally, $ N_i$ are the corresponding particle numbers of these species.  Recall that the entropy $S$ is a state function of $U,V$ and of all the  $N_i$: $S=S(U,V, N_1.N_2,...)$. It is convenient to express the above law in terms of  the energy density, $\rho=U/V$, entropy density, $s=S/V$ and number density for each  ith species, $n_i=N_i/V$.  Expanding Eq.\,(\ref{eq:FirstLawThermo}) in terms of them, we find
\begin{equation}\label{eq:FirstLawThermobis1}
(Ts-p-\rho+\sum_i\mu_i n_i)dV+(Tds-d\rho+\sum_i\mu_i dn_i)V=0\,.
\end{equation}
This relation can now be first applied to an arbitrary region of the system with constant volume, $V=$const.  It immediately follows from \eqref{eq:FirstLawThermobis1} that  $Tds-d\rho+\sum_i\mu_i dn_i=0$.  Obviously, this differential relation does not depend on the volume since it comprises, on the one hand,  quantities such as energy density, entropy density and the number densities, and on the other it involves the intensive parameters $T$ and $\mu_i$. Thus, we may now use such differential relation back on \eqref{eq:FirstLawThermobis1}, as  the latter is valid both for the entire system or any of its parts\,\cite{Rubakov:2018}.  In this way the following well-known relation ensues immediately:
\begin{equation}\label{eq:entropydensity}
  s=\frac{\rho+p-\sum_i\mu_i n_i}{T}\,.
\end{equation}
An alternative, perhaps more conventional, derivation of this important relation can be obtained by invoking the integrability condition of the full differential form $dS$, i.e. $\partial^2S/\partial T\partial V=\partial^2S/\partial V\partial T$\,\cite{KolbTurner:1990};  the interested reader may  check the Appendix A for details. }

In the cosmological context,  $V$ will be in general  the physical volume of a considered part of the universe at a time where the scale factor is $a$. Thus,  we have  $V=a^3 L^3$, where  $L^3$ is the coordinate volume at present\,\footnote{We take  $a$  dimensionless and normalized as $a(t_0)=1$ at the present value of the comic time, $t=t_0$.}. Usually, for simplicity, we will assume the coordinate volume $L^3=1$ and then $V$ just coincides with the comoving volume $a^3$. Thus, $S=s a^3$ is the total comoving entropy in the comoving volume and $s=n\sigma$ is the aforementioned entropy density, i.e. the $0th$ component of the entropy flow vector (\ref{entropydefination}).  
For the comoving volume $V=a^3$, Eq.\,(\ref{eq:FirstLawThermo}) can be written in the cosmological contest as follows:
\begin{equation}\label{eq:FirstLawThermo2}
Td(s a^3)-d(\rho a^3)-p da^3+\sum_i\mu_i d(n_i a^3)=0.
\end{equation}

Hereafter, for simplicity, we will discuss the case of one species of particles only and we shall therefore omit the sum over species (as we already did in the initial considerations of this section).
In this notation, and taking into account that $s=n\sigma$, Eq.\,(\ref{eq:entropydensity}) implies that the chemical potential of that particular species  is given by
\begin{eqnarray}
\mu=\frac{\rho+p}n-T\sigma.\label{chemicalpone}
\end{eqnarray}
If $\sigma$ is assumed constant, i.e. $\dot{\sigma}=0$ (see further discussions below), it is easy to combine equations \eqref{eq:FirstLawThermo2} and \eqref{chemicalpone} to obtain (for one component) the following differential law:
\begin{equation}\label{eq:FirstLawOpenSystem}
d(\rho a^3)+p d a^3=\frac{\rho+p}{n}d(n a^3)\,.
\end{equation}
This law is sometimes called the First Law of thermodynamics generalized for open systems\,\cite{Prigogine:1986zz,Prigogine:1988zz,Prigogine:1989zz}. In fact, the cosmic system under study is open since the evolution of $\rho$ is not only affected by the dilution produced by the universe's expansion but also by the creation (or disappearance) of new particles in the comoving volume. The above equation can be conveniently cast as
\begin{equation}\label{eq:conservationlawMatter1}
\rho'+\frac{3}{a}(p+\rho)=\left(\frac{\rho+p}{N}\right)\,\frac{dN}{da}\,,
\end{equation}
where prime denotes $d/da$. As always, $N=n a^3$ stands for the total number of particles in the comoving volume.
Notice also that because $\rho+p$ is the enthalpy per unit volume\,\cite{BookCallen1960}, ${\cal H}/V$, the coefficient of the differential in the \textit{r.h.s.} of Eq.\,\eqref{eq:FirstLawOpenSystem} is just the specific enthalpy (i.e. the enthalpy associated to a single particle): $(\rho+p)/n={\cal H}/N$. The presence of the enthalpy here is quite natural if we take into account that it acts as potential for work for systems of constant pressure, these systems being generally open systems for which \eqref{eq:FirstLawOpenSystem} holds.  For a closed system with a fixed number of particles, however, the \textit{r.h.s.} of equations\,\eqref{eq:FirstLawOpenSystem} and \eqref{eq:conservationlawMatter1} vanishes since $d(n a^3)=dN=0$, and one then recovers the First Law for closed systems:
\begin{equation}\label{eq:conservationlawMatter2}
\rho'+\frac{3}{a}(p+\rho)=0\,.
\end{equation}

Another important differential form in thermodynamics is the Gibbs-Duhem equation\,\cite{BookCallen1960}: $S dT-Vdp+N d\mu=0$. The latter follows at once from differentiating Euler's relation $U=TS-pV+\mu N$ and using the First Law (\ref{eq:FirstLawThermo}). Recall that the mentioned Euler's relation is a consequence of the fact that the total entropy is a homogeneous function of degree one in its natural variables $S=S(U,V,N)$\,\cite{BookCallen1960}. The Gibbs-Duhem equation shows that the three intensive variables $T, p$ and $\mu$ are not independent; if we know two of them, the value of the third can be determined from such differential expression.
We are now ready for an important step, which is to derive the relation between entropy flow and particle production. To this end, we combine the Gibbs-Duhem equation with Eq.~(\ref{chemicalpone}) and its own differential form.  A straightforward calculation leads to
\begin{eqnarray}
n T d\sigma=d\rho-\frac{\rho+p}{n} d n.\label{gibbs}
\end{eqnarray}
Dividing it out by $T dt$,  one gets
\begin{eqnarray}
n \dot \sigma=\frac{\dot\rho}{T}-\frac{(\rho+p)}{T}\frac{\dot n}{n}\,.\label{gibbs1}
\end{eqnarray}
Using (\ref{balanceLaw}) for $\dot{n}/n$  and inserting (\ref{gibbs1})
in (\ref{divergenceSflow}), we find:
\begin{eqnarray}
\nabla_\alpha s^\alpha=\frac{\dot\rho}{T}-\frac{(\rho+p)}{T}(\Gamma-\theta)+\sigma  n {\Gamma}\,.
\label{divergenceS2}
\end{eqnarray}

\subsection{Production pressure and bulk viscosity}

Another important ingredient which we need to discuss is the notion of production pressure and bulk viscosity in the context of particle production.  If there is no particle production, the usual local conservation law in cosmology is $\nabla_\mu T^{\mu\nu}=0$, where $T^{\mu\nu}$ is the energy-momentum tensor for matter. If all components making up the universe can be considered as perfect fluids,  the energy-momentum tensor of all matter in the FLRW metric takes on the form
\begin{eqnarray}
T_{\mu\nu}=(\rho+p)u_\mu u_\nu+p g_{\mu\nu}.\label{emt}
\end{eqnarray}
The parameters $p$ and $\rho$ are the total (proper) pressure and energy density of the matter fluids, respectively.
Using the above perfect fluid form and the FLRW metric (\ref{eq:FLRWmetric}), the $\nu=0$ component of the conservation law $\nabla_\mu T^{\mu\nu}=0$ just reads
\begin{equation}\label{eq:conservationlawMatter}
\dot\rho+3H(p+\rho)=0\,.
\end{equation}
This equation is just an alternative form of Eq.\,\eqref{eq:conservationlawMatter2} after we trade the cosmic time for the scale factor in the process of differentiation (recall that $d/dt=a H d/da$), as could be expected.
In the presence of  particle production, however,  it is customary to associate a production pressure, $p_c$, to the creation of new particles. As a result the above conservation law gets modified as follows:
\begin{eqnarray}
\dot\rho+3H(p+\rho+p_c)=0.\label{conservation11}
\end{eqnarray}
Combining this  equation with (\ref{chemicalpone}) for the chemical potential, we can rewrite the relation between the entropy flow and particle production, Eq.\,(\ref{divergenceS2}), in the more compact form:
\begin{eqnarray}
\nabla_\alpha s^\alpha=-\frac{p_c\theta}{T}-\frac{\mu \psi}{T}=-3 H\frac{p_c}{T}-\frac{\mu n \Gamma}{T}\,,
\label{divergence3}
\end{eqnarray}
where we recall that $\theta=3H$ is the expansion of the cosmological fluid in FLRW spacetime. This formula has been introduced in~\cite{Calvao:1991wg} and applied to different contexts.

Particle production can mimic a fluid having bulk viscosity pressure $\Pi=-\zeta\theta$, where $\zeta$ is the bulk viscosity parameter (having dimension 3 of energy, in natural units)\,\cite{RoyMaartens1996}. If there are no other dissipative processes except bulk viscosity, we have
\begin{eqnarray}
\dot\rho+\theta(p+\rho+\Pi)=0\,,\label{conservationpi}
\end{eqnarray}
where the entropy production is fully determined by $\Pi$ through\,\cite{RoyMaartens1996}
\begin{equation}\label{nablaSbulk}
  \nabla_\alpha s^\alpha=\frac{\Pi^2}{\zeta T}\,.
\end{equation}
Clearly, Eqs.~(\ref{conservation11}) and (\ref{conservationpi}) are very similar since in the cosmological context  $\theta=3H$. We can see that particle creation can be conceived as a form of bulk viscosity. Indeed, if $p_c$ is to satisfy also (\ref{nablaSbulk}), i.e. $ \nabla_\alpha s^\alpha={p_c^2}/{(\zeta T)}$, then from  (\ref{divergence3}) we find
$p_c^2+\zeta p_c\theta=-\zeta n \mu \Gamma$.
Thus, for $\Gamma=0$ we recover the relation  $p_c=-\zeta\theta$, which is alike to the one satisfied by $\Pi$ in the absence of particle production, whereas  for $\Gamma\neq 0$  particle production leads to a generalized relation between $p_c$, bulk viscosity and the fluid expansion.

It is also customary to parameterize the  production of particles in an expanding universe  in terms of an anomalous conservation law as follows\,\cite{Lima1996}:
\begin{eqnarray}
\dot{\rho}+3H (\rho+p)=\beta n\Gamma\,,\label{AnomalousConservLaw}
\end{eqnarray}
where $\beta>0$ is a dynamical quantity with dimensions of energy, and $\Gamma$ is the particle production rate mentioned above.  For $\Gamma=0$ we recover the standard local conservation law (\ref{eq:conservationlawMatter}). From equations (\ref{conservation11}) and (\ref{AnomalousConservLaw}) we  find
\begin{eqnarray}
p_c=-\beta \frac{n\Gamma}{3H}\,.
\label{Ansatzpc}
\end{eqnarray}
The production pressure  vanishes in the absence of particle production, as expected.  On the other hand, for an expanding universe with particle production,
we have $H>0$ and $\Gamma>0$ and hence the production pressure satisfies $p_c<0$ ($\beta>0$).
The induced  effective negative pressure is the reason why viscous fluids  can be conceived as a mechanism for acceleration
even without dark energy~\cite{Ford:1986sy,Traschen:1990sw,Abramo:1996ip,Gunzig:1997tk,
Peebles:1998qn,Zimdahl96,Allahverdi:2010xz}.  Moreover the bulk viscosity suppresses the growth of structures since the induced negative pressure works against the
gravitational collapse. Similarly, the shear viscosity \cite{RoyMaartens1996} refrains the velocity perturbations from increasing and this also stops the growth, but we shall not address this effect here.

Taking Eq.~(\ref{Ansatzpc}) into Eq.~(\ref{divergence3}) we obtain
\begin{eqnarray}
\nabla_\alpha s^\alpha&=&\frac{n\Gamma}{T}(\beta-\mu)\nonumber\\
&=&n\Gamma\sigma+\left(\beta-\frac{\rho+p}{n}\right)\frac{n\Gamma}{T},
\label{energyconservation4}
\end{eqnarray}
where we used Eq.~(\ref{chemicalpone}) to get the second equality. For $\Gamma>0$ (particle production) and $\beta>\mu$, the entropy of particles is growing. If $\beta<\mu$, we need $\Gamma<0$, i.e. particle annihilation, in order to preserve the Second Law of Thermodynamics.

\subsection{Entropy evolution and Second Law in the expanding universe}

Finally, we consider  the evolution of the total entropy.  If $N$ is the total number of particles of a given sort in a comoving volume $a^3$, i.e. $N=n a^3$,  the total entropy in the comoving volume  is just $N$ times the specific entropy of particles: $S=N\sigma$. Therefore
\begin{eqnarray}
\frac{\dot S}{S}=\frac{\dot N}{N}+\frac{\dot\sigma}{\sigma}
=\Gamma+\frac{\dot\sigma}{\sigma}\,,\label{dotentropy1}
\end{eqnarray}
where in the last step we have used (\ref{GammaAlternative}).
It is convenient to rewrite the general equation (\ref{dotentropy1}) as follows:
\begin{eqnarray}
\frac{dS}{dt}=S\left(\frac{\dot\sigma}{\sigma}+\Gamma\right)=N\dot{\sigma}
+N\sigma\Gamma.\label{dotentropy2}
\end{eqnarray}
When $\dot\sigma$ is nonvanishing the evolution of the entropy can be compatible with the presence of particle annihilation   ($\Gamma<0$) since we can still preserve the Second Law  ($\frac{dS}{dt}>0$)  provided the rate of change of $\sigma$ with the expansion is large enough. However, in the frequent special case  $\dot\sigma=0$ (i.e. constant specific entropy for each particle) the evolution of the entropy is fully controlled by particle production:
\begin{equation}\label{eq:sigmaconstant}
\frac{\dot S}{S}=\frac{\dot N}{N}=\Gamma\,,
\end{equation}
and then the Second Law requires  $\Gamma>0$.  Let us now recall that it is precisely when $\dot\sigma=0$ that equation \eqref{eq:conservationlawMatter1} is fulfilled. If we express such equation in terms of the cosmic time, it is easy to find
\begin{equation}\label{eq:conservationlawMatter3}
\dot\rho+3 H (p+\rho)=(\rho+p)\,\frac{\dot{N}}{N}= \Gamma\,(\rho+p)\,,
\end{equation}
where once more we have used (\ref{GammaAlternative}). For $\Gamma=0$ we naturally retrieve Eq.\,\eqref{eq:conservationlawMatter}, as it should be expected. Furthermore,  on comparing \eqref{eq:conservationlawMatter3} with \eqref{AnomalousConservLaw} we find that
\begin{equation}\label{betaequation}
  \beta=\frac{\rho+p}{n}\,.
\end{equation}
Thus, from the foregoing discussion it follows that when $\dot\sigma=0$ the parameter $\beta$ becomes equal to the the specific enthalpy. We shall further discuss this relation when we consider vacuum decay into matter in the next section.

Equation (\ref{dotentropy2}) is the counterpart of (\ref{divergenceSflow}) for the total entropy evolution law. The \textit{r.h.s.} of equations  (\ref{divergenceSflow}) and (\ref{dotentropy2}) differ by a factor $V=a^3$. Thus,  $dS/dt=a^3\nabla^\alpha s_\alpha$, as noted already in Eq.\,\eqref{eq:nablas2} above, and we can equivalently express the Second Law of Thermodynamics either as $\frac{dS}{dt}>0$ or as  $\nabla_{\alpha}s^{\alpha}>0$.  But this is only one of the conditions to be satisfied by the entropy function. In fact, the basic thermodynamic conditions associated to the Second Law are two:  i) the entropy must increases with the expansion; and  ii) its increase must wane with time until the system (in this case the universe) eventually achieves a final state of thermodynamic equilibrium. Therefore, we actually need the following two conditions\footnote{For simplicity, we restrict to the case of a single variable $\xi$, as this will be sufficient for our purposes. In general, the conditions insuring a path to stable equilibrium are expressed in terms of the first and second order differentials, namely $dS\geq0$ and $d^2S<0$, in whatever number of variables\,\cite{BookCallen1960}. }:
\begin{equation}\label{eq:TwoConditions}
\frac{dS}{d\xi}\geq0 \ \ \ \ \ \ \ \  \frac{d^2S}{d\xi^2}<0\,.
\end{equation}
While the first condition is usually associated to the Second Law, the second one (or stability condition) is also necessary and is dubbed the Law of Thermodynamic
Equilibrium (LTE). In its absence the entropy could raise in an uncontrolled way. We will see the importance of the LTE throughout our discussions.
In cosmology,  the differentiation is to be performed with respect to some relevant variable $\xi$ associated to the increasing expansion of the universe, for example the cosmic time or the scale factor. The equilibrium state is finally attained when the first derivative is zero and the second remains negative (which defines, of course, the concavity condition around the maximum of the function  $S$).  The situation  $dS/d\xi=0$ and $d^2S/d\xi^2<0$ therefore corresponds to the state of stable equilibrium at a maximum entropy.  We can check it explicitly from Eq.\,(\ref{eq:FirstLawThermo2}), when the variable $\xi$ is taken to be the cosmic time of the FLRW metric. Dividing out Eq.\,(\ref{eq:FirstLawThermo2}) by $dt$, it becomes
\begin{equation}\label{eq:Conservation entropy}
  T\frac{d(s a^3)}{dt}=\frac{d(\rho a^3)}{dt}+ p\frac{d a^3}{dt}-\sum_i {\mu_i}\frac{d(n_i a^3)}{dt}=a^3\left[\dot{\rho}+3H(\rho+p)\right]-\sum_i{\mu_i}\frac{d(n_i a^3)}{dt}\,.
\end{equation}
Assuming local covariant conservation law of matter, Eq.\,\eqref{eq:conservationlawMatter}, the above equation boils down to
\begin{equation}\label{eq:Conservation entropy2}
  \frac{dS}{dt}=-\sum_i\frac{\mu_i}{T}\frac{d(n_i a^3)}{dt}\,.
\end{equation}
This equation shows that the total entropy in the comoving volume, $S=s a^3$,  will not be conserved if there is creation or destruction of particles. A path to equilibrium requires $dS/dt\geq 0$ and $d^2S/dt^2< 0$.
However, for quasi-static processes around equilibrium we have  $d(n_i a^3)=dN_i=0$ for each species of particles since the chemical potentials correspond to conserved quantities in the state of equilibrium (e.g. charge, hypercharge, baryon or lepton number  etc, which cannot increase nor decrease at this point), and  on the other hand $\mu_i=0$ for those species not carrying any conserved quantum number (such as photons or Majorana particles). Therefore, for these processes  the last term of \,Eq.\,(\ref{eq:Conservation entropy2}) vanishes and we achieve  $dS/dt=0$; and stability requires also $d^2S/dt^2<0$. In other words, in such conditions we may consider that the entropy is conserved at a maximum value in a comoving volume.  The evolution  of the universe in certain stages can be considered to satisfy such entropy conservation law, typically for sufficiently large temperatures where the interaction rates of particles are larger than the expansion rate\,\cite{KolbTurner:1990}. The above equations, however, must be further modified in the presence of vacuum dynamics since the vacuum can be itself a source  of new particles or a sink of them.  This will be dealt with in the next section.

\mysection{Dynamical vacuum models}
\label{secDVM}
We have reviewed some basic knowledge of entropy and particle production in the cosmological context. With the previous discussion, we can apply these generic results to cosmological models where the vacuum is not a mere spectator. The possibility that the vacuum energy density is an evolving quantity  can be considered a natural assumption in an expanding universe. One might even deem that the more standard assumption $\rL=$const. (which is the usual trademark  in the concordance $\CC$CDM model)  is an oversimplification in the context of an expanding cosmological background. Let us emphasize that Occam's razor is preferred only if it works better, but sometimes additional assumptions can be more efficient.   The idea of dynamical vacuum energy can be helpful for the study of our speeding up universe, and it proves competitive enough in the description of the overall cosmological data~\cite{Gomez-Valent:2014rxa,Sola:2015wwa,Sola:2016jky,Sola:2017znb,
Sola:2016hnq,Sola:2016ecz,Sola:2017jbl,Rezaei2019}.  For this reason we shall focus on the thermodynamical aspects of   the class of the dynamical vacuum models (DVMs) and particulary on the subclass of the running vacuum models (RVMs), which we will introduce in more detail in Sec.\,\ref{RVM}.

\subsection{Einstein's equations with dynamical vacuum}

Let us start with Einstein's equations, which  can be written  as follows:
\begin{eqnarray}
G_{\mu\nu}+\Lambda g_{\mu\nu}=\kappa^2 T_{\mu\nu}\,.\label{einteinequation}
\end{eqnarray}
Here, $G_{\mu\nu}=R_{\mu\nu}-(1/2) g_{\mu\nu} R$ is the Einstein tensor, and we have defined $\kappa^2\equiv {8\pi G}$ with  $G$  Newton's constant. In addition,  $\Lambda$ is the cosmological term, which has no obvious  relation with matter. Expressed as a vacuum  energy density, the quantity $\rL=\CC/\kappa^2$  is of order of the current critical density $\rco$ (specifically $\rL\simeq 0.7\rco$\,\cite{SNIa,Planck2015-18}) and hence very small as compared to any typical value in particle physics. One finds $\rL\sim 10^{-47}$ GeV$^4$, which  is much smaller than $m^4$ for any known particle mass except for a light neutrino in the millielectronvolt range\,\cite{JSPRev2013}.  We also know observationally that $\rL$ is approximately constant, but we cannot exclude a certain evolution with the cosmic expansion.  Such evolution is perfectly allowed by the
Cosmological Principle, which expresses the homogeneity and isotropy  of the large scale observations.  Studies in recent years indicate the possibility that the fundamental ``constants'' of Nature may evolve mildly with the expansion (see e.g.~\cite{Uzan:2011,Chiba:2011,Magueijo:2003}). It has been suggested that a possible explanation could be an exchange of energy between matter and vacuum\,\cite{FritzschSola2012}.  In this work we will assume such possibility, and more specifically that the vacuum energy density $\rL$ is actually  ``running'' with the cosmic expansion rate $H$\cite{JSPRev2013}, although we will keep $G=$const. for the present work\footnote{See Refs.~\cite{Sola:2015wwa,Sola:2016jky,Nesseris2017} for some recent studies considering the possibility to improve the fit to the current  cosmological data in models  where $G$ evolves with time.}.

In the context of DVMs, the Einstein equation can be formally written as in (\ref{einteinequation}) but admitting the possibility that $\CC=\CC(t)$, and hence $\rL=\rL(t)$, is evolving with the cosmic time:
\begin{eqnarray}
G_{\mu\nu}=\kappa^2\left(T_{\mu\nu}-g_{\mu\nu} \rL(t)\right)\,.\label{einteinequationrL}
\end{eqnarray}
From the Bianchi identity, $\nabla^\mu G_{\mu\nu}=0$, it follows that the matter energy-momentum tensor $T_{\mu\nu}$ is constrained to satisfy
\begin{eqnarray}
\nabla^\mu T_{\mu\nu}=g_{\mu\nu}\nabla^\mu\rL(t)\,,\label{energyconservation}
\end{eqnarray}
and hence it cannot be conserved if $\rL(t)$ is an evolving cosmic variable.  Using the perfect fluid form (\ref{emt}) and
taking the $\nu=0$ component of the above equation in the FLRW metric, we find
\begin{eqnarray}
\dot\rho+3H(\rho+p)=-\drL(t)\,.\label{energyconservation2}
\end{eqnarray}
This equation expresses that the total energy density and pressure of matter (both relativistic and nonrelativistic), $\rho$ and $p$, could be in general interacting with the vacuum components  $\rL$ and $p_\CC$. Recall that the latter satisfy the equation of state (EoS)
 \begin{equation}\label{eq:EoSvacuum}
p_\CC(t)=-\rL(t)\,.
\end{equation}
We should emphasize at this point that owing to its definition we assign the same EoS to vacuum whether static or dynamical: $w_\CC=p_\CC/\rL=-1$. Equation \,\eqref{energyconservation2} generalizes (\ref{eq:conservationlawMatter}) in the presence of vacuum energy in interaction with matter. Being the combined system of matter and vacuum a closed system in interaction, it must also preserve as a whole the local covariant conservation law of energy. We can check that if we use the total energy density and pressure of matter and vacuum, namely $\rho_T=\rho+\rL$ and $p_T=p+p_\CC$, in equation (\ref{eq:conservationlawMatter}) -- i.e. if we replace $\rho\to\rho_T$ and $p\to p_T$ in it --  we recover Eq.\,(\ref{energyconservation2}).

The corresponding form of Einstein's equations for the spatially flat FLRW metric are the well-known pair of Friedmann's equations for the total energy density and pressure:
\begin{eqnarray}
&&3H^2=\kappa^2\rho_T=\kappa^2\sum_{i=m,r,\CC}\rho_i=\kappa^2(\rmm+\rr+\rL)\,,
\label{eq:FriedmannEq}\\
&&3H^2+2\dot{H}=-\kappa^2 p_T=-\kappa^2\sum_{i=m,r,\CC}p_i=-\kappa^2 (p_m+p_r+\pL)=\kappa^2(\rL-\frac13\,\rr)\label{eq:PressureEq}\,,\,
\end{eqnarray}
where the sum is over all the components of the cosmic fluid: $i=m,r,\CC$,  i.e. non-relativistic matter, radiation and vacuum, with $w_i=p_i/\rho_i$ the EoS parameters for each component. Notice that the total matter energy density (relativistic plus non-relativistic) is $\rho=\rr+\rmm$, where $\rmm=\rb+\rcdm$ is the total non-relativistic part, the latter involving the contributions from baryons and CDM, both with vanishing pressure. The relativistic part (or radiation) and the vacuum part, instead, are characterized by the EoS's $p_r=\rho_r/3$ and (\ref{eq:EoSvacuum}), respectively. One can readily check that  the local conservation law (\ref{energyconservation2}) can be recovered directly from the previous equations (\ref{eq:FriedmannEq}) and (\ref{eq:PressureEq}), and can be expressed in a compact manner as follows:
\begin{equation}\label{eq:GeneralCL}
\dot\rho_T+3H(\rho_T+p_T)=\sum_{i=m,r,\CC}\dot{\rho}_{i}+3 H(1+w_i)\rho_i=0\,.
\end{equation}
Such equation  is thereby not independent from the above pair of Friedmann's equations, but it is very useful as it allows to express the local covariant conservation of all the components of the cosmic fluid. It splits into the corresponding conservation equations for each fluid component only if they are all free, i.e. non-interacting. But this need not to be the case for some of the components.

\subsection{Interaction of matter with vacuum}

Equation\,(\ref {eq:GeneralCL}) can  be put in the form
\begin{equation}\label{eq:Qequations}
\dot{\rho}+3H\rho=Q\,,\ \ \ \ \ \ \, \dot\rho_{\CC}=-{Q}\,.
\end{equation}
which is just a convenient way of expressing Eq.\,(\ref{energyconservation2}) in which an interaction source $Q$ is explicitly introduced, but at this point is still unspecified.
One  frequently assumes that the density of baryonic matter  is conserved ($\dot\rho_b+3H\rho_b=0$) after it was first created in the early universe\,\footnote{A mechanism for baryogenesis through leptogenesis has been proposed within the RVM context, see \cite{Anomaly2019a} and references therein, as well as the additional comments in the next section.}.  Furthermore,  since radiation is subdominant at present it means that if there is to be a nontrivial energy exchange at present it is usually assumed to occur between cold DM and vacuum only. Under this assumption, we can just replace $\rho\to\rho_{cdm}$ in \eqref{eq:Qequations}. However, the general thermodynamic  considerations under study will not depend on these microphysical details, and so we  need not make very specific assumptions of this kind here.
Different Ans\"atze for $Q$ used in the literature are usually proportional to the Hubble rate times a linear combination of energy densities of various sorts:
\begin{eqnarray}
Q=3 H\sum_i\nu_i\rho_i\,,\label{Sources}
\end{eqnarray}
where $\nu_i$ are dimensionless coefficients and the factor $3$ is for convenience. The energy densities can be of CDM, baryons, radiation  or even vacuum energy density, depending on the specific assumptions made, the coefficient being zero if the given species does not interact;  see e.g. \cite{Sola:2017jbl} and references therein for particular instances, which have been thoroughly studied in the literature and compared with observations.

The relationship between particle production rate and the decaying vacuum $\Lambda(t)$ can be obtained from  the evolution equations of the specific entropy ($\sigma$) and temperature ($T$) of the created particles.  The vacuum is assumed to have no chemical potential ($\mu_\CC=0$) and hence its EoS \eqref{eq:EoSvacuum}  implies that it has no entropy\,\cite{Lima1996,Graef:2013iia}.  This is obvious from (\ref{eq:entropydensity}) or (\ref{chemicalpone}) if we call upon \eqref{eq:EoSvacuum}.
Assuming that the particle production from vacuum is adiabatic, such that some basic thermodynamic equilibrium relations are preserved, one finds that the quantity $\beta$ introduced previously in Eq.\,(\ref{AnomalousConservLaw}) becomes determined as follows (see~\cite{Calvao:1991wg,Lima1996} for details):
\begin{equation}\label{betaequation2}
  \beta=\frac{\rho+p}{n}\,.
\end{equation}
Recall from Sec.\,\ref{sec2} that $\rho+p\equiv {\cal H}/V$ is the enthalpy per unit volume, so we have $\beta={\cal H}/N$; and hence under the mentioned thermodynamical conditions in cosmology, $\beta$ becomes equal to the the specific enthalpy.
Let us now show that the above relation implies  $\dot \sigma=0$. This follows immediately from equating the two alternative expressions that we have found in Sec.~\ref{sec2}  for the entropy flow, namely equations \eqref{divergenceSflow}   and  \eqref{energyconservation4}. We obtain
\begin{eqnarray}\label{5}
n\dot{\sigma}=\left(\beta-\frac{\rho+p}{n}\right)\frac{n\Gamma}{T}\,.
\end{eqnarray}
So indeed the fulfillment of \eqref{betaequation2} automatically implies  $\dot{\sigma}=0$. Thus, when Eq.\,(\ref{betaequation2}) is fulfilled the entropy change can only be due to the change in the number of particles, see Eq.\,(\ref{eq:sigmaconstant}). Notice that Eq.\,\eqref{betaequation2} is the same as Eq.\,\eqref{betaequation}, except that here the relation $\dot\sigma=0$ has been inferred rather than assumed. Such alternative derivation ~\cite{Calvao:1991wg,Lima1996} gives a new physical insight on the meaning of the condition $\dot{\sigma}=0$, as it shows that this relation is automatically implied whenever the particles originated from the decaying vacuum are created immediately in equilibrium with the already existing ones, i.e. when no finite time is needed for thermalization. This  will be assumed throughout our study unless stated otherwise\,\footnote{If one wishes to study an evolving $\sigma$ over time while ensuring that \eqref{betaequation2} remains  true, it is necessary to introduce an extra term into the conservation equation of matter~\cite{Calvao:1991wg,Lima1996}.}.

As a consequence of the relation \eqref{betaequation2} we can now derive an important formula, which follows from Eqs.\,(\ref{AnomalousConservLaw}) and (\ref{energyconservation2}):
\begin{eqnarray}
\Gamma=\frac{-\drL(t)}{n\,\beta}=\frac{-\drL(t)}{\rho+p}=\frac{-\drL(t)}{(1+w)\rho}\,,
\label{relation1}
\end{eqnarray}
where $w=p/\rho$ is the EoS of matter, relativistic ($w_r=1/3$) or nonrelativistic ($w_m=0$). This result is reasonable: if the entropy per particle is constant ($\dot\sigma=0$), then, when $\rL(t)$ decays ($\drL(t)<0$),  the energy of vacuum is fully invested in the creation  of new particles ($\Gamma>0$),  whereas if $\drL(t)>0$ the particles disappear into vacuum ($\Gamma<0$).  We shall see that the above relation plays an important role in our discussion.

We conclude this section with following remark. In the $\CC$CDM, the vacuum cannot decay into particles, so we have $\Gamma=0$.  In this case, Eq.\,\eqref{5} does not apply since it is based on \eqref{energyconservation4}, which does not apply either since it assumes nonvanishing particle production.  However, the primary relation\,\eqref{divergenceSflow}  still holds and since $\Gamma=0$ we find  $\nabla_\alpha s^\alpha=n \dot{\sigma}$, and so in this case the evolution of the entropy is fully controlled by $\dot{\sigma}$.  On comparing the previous equation with \eqref{eq:nablas2} we find that $n \dot{\sigma}=(1/a^3) dS/dt$, or
\begin{equation}\label{eq:dotsigma}
  \dot{\sigma}=\frac{1}{n a^3}\,\frac{dS}{dt}=\frac{\dot S}{N}\,,
\end{equation}
which is perfectly consistent with $S=N\sigma$ since $N$ (the total number of particles in the comoving volume) is conserved in this case. In this particular situation, the Second Law of Thermodynamics obviously implies $\dot{\sigma}>0$ unless the universe already reached equilibrium.

\mysection{Running Vacuum Models}
\label{RVM}
In this section, we apply the thermodynamic framework discussed so far to some specific DVMs as this may help to have a better understanding of these models. The particular case of the running vacuum model (RVM)  provides a competitive fit to the overall set of cosmological observations as compared to the concordance model of cosmology, i.e.  the $\CC$CDM, see e.g. ~\cite{Gomez-Valent:2014rxa,Sola:2015wwa,Sola:2016jky,Sola:2017znb,
Sola:2016hnq,Sola:2016ecz,Sola:2017jbl,Rezaei2019}. The RVM can also be extended such that it can also encompass the physics of inflation and its transition to the standard radiation epoch, see\,\cite{Lima:2012mu,Perico:2013mna} and \cite{Sola:2015rra,GRF2015}.

\subsection{Unified model of vacuum energy density}
The RVM is well described in the aforesaid references (see also\,\cite{JSPRev2013,Sola:2015rra} for a review)  and here we limit ourselves to point out the basic formulas. The model is based on a renormalization group equation which governs  the ``running'' of the vacuum energy density $\rL$ as a function of the Hubble rate $H$.  Up to ${\cal O}(H^4)$ it reads \begin{equation}\label{runningH}
\frac{d\, \rho_\Lambda (H)}{d\, {\rm ln}H^2 } = \frac{1}{(4\pi)^2}
\sum_i \Big[a_i M_i^2 H^2 + b_i H^4 \Big]\,,
\end{equation}
where the coefficients $a_i, b_i$ are dimensionless
and receive contributions from
loop corrections of boson  (B) and fermion (F)  matter fields with different
masses $M_i$.
Integration leads to the vacuum energy density\,\footnote{The general case for higher order powers of the Hubble rate,  $H^{n+2}\, (n\geqslant 2)$, is studied in the Appendix B.}
\begin{equation}\label{eq:rhoLambdaunified}
\rho_{\Lambda}(H) = \frac{\Lambda(H)}{\kappa^2}=
\frac{3}{\kappa^2}\left(c_0 + \nu H^{2} + \alpha
\frac{H^{4}}{H_{I}^{2}}\right)\,.
\end{equation}
Once more we have denoted  $\kappa^{2}\equiv8\pi G$, and
$c_0$ is an integration constant (with dimension $+2$ in
natural units, i.e. energy squared) which together with the dimensionless coefficient $\nu$ can be constrained
from the cosmological data.
The latter is formally given in QFT as
\begin{equation}\label{eq:loopcoeff}
\nu=\frac{1}{48\pi^2}\, \sum_{i=F,B} a_i\frac{M_i^2}{M_{\rm
Pl}^2}\,.
\end{equation}
Notice that $M_{\rm Pl}=1/\kappa=M_P/\sqrt{8\pi}$ is the reduced Planck mass, where   $M_P=1/\sqrt{G}$ is the usual Planck mass (in natural units).  Clearly, only the heavy particle masses $M_i$ from the Grand Unified Theories (GUT's) can provide a non-negligible contribution\,\cite{JSPRev2013}. On the other hand, the coefficient controlling the higher order power in $H$ reads
\begin{equation}\label{eq:alphaloopcoeff}
\alpha=\frac{1}{96\pi^2}\, \frac{H_I^2}{M_{\rm Pl}^2}\sum_{i=F,B}
b_i\,.
\end{equation}
The dimensionless coefficients $(\nu,\alpha)$ play the role of
one-loop beta-functions (at the respective low and high energy
regimes during the universe expansion). They are expected to be naturally small because $M_i^2\ll
M_{\rm Pl}^2$ for all the particles, even for the heavy fields of a
typical GUT, where $\nu$ lies
in the range $|\nu|=10^{-6}-10^{-3}$ \cite{Fossil07}, while
$\alpha$ is also small ($|\alpha|\ll 1$),
since the Hubble rate  at the scale of inflation, $H_I$, is certainly below the Planck
scale.  At low energies, well after the inflationary period,  the dynamical properties of the vacuum are fully controlled by $\nu$. The latter thereby plays the role of the observational running vacuum parameter for the current universe, see Sec.~\ref{RVMCurrentUniverse}.  Needless to say, for $\nu=\alpha=0$ we recover the standard $\CC$CDM model with no dynamical evolution of the vacuum.

Remarkably, the possible dynamical evolution of the vacuum and hence the departure from the rigid $\CC=$const. picture of the $\CC$CDM can be motivated on fundamental grounds. Apart from its possible connection to QFT in curved spacetime in different frameworks\,\cite{ShapSol,Fossil07,RVM-SUGRA} -- see also \,\cite{JSPRev2013} for a review --  it has recently  been shown that the presence of the term  $H^4$ in the effective expression of the vacuum energy density, Eq.\,(\ref{eq:rhoLambdaunified}), can be the generic result of the low-energy effective action based on the bosonic gravitational multiplet of string theory, see \cite{Anomaly2019a}. The unavoidable presence of the (CP-violating) gravitational Chern-Simons term associated with that action turns out to lead to an effective $\sim H^4$ behavior when averaged over the inflationary spacetime, in the presence of primordial gravitational waves. This higher order term triggers inflation within the context of the RVM, see Sec.~\ref{RVMearlyUniverse}. In the early universe, before and during inflation, it is assumed that only fields from the gravitational multiplet of the string exist, which implies that the relevant bosonic part of the effective action pertinent to the dynamics of the inflationary period is given by\,\cite{Anomaly2019a}
\begin{equation}\label{gCS}
S^{\rm eff}_B =\; \int d^{4}x\sqrt{-g}\Big[ -\dfrac{1}{2\kappa^{2}}\, R + \frac{1}{2}\, \partial_\mu b \, \partial^\mu b +   \sqrt{\frac{2}{3}}\,
\frac{\alpha^\prime}{96 \, \kappa} \, b(x) \, R_{\mu\nu\rho\sigma}\, \widetilde R^{\mu\nu\rho\sigma} + \dots \Big]\,.
\end{equation}
It involves the usual Hilbert-Einstein term and the Kalb-Ramond (KR) axion field,  $b(x)$, which  is coupled to the gravitational Chern-Simons topological density through the string tension $\alpha^\prime$. As indicated, such topological term  when averaged over the de Sitter spacetime produces an effective  contribution to the vacuum energy density of the form  $\sim H^4$, similar to the last term in Eq.\,(\ref{eq:rhoLambdaunified}). Let us also remark that, in such a context, one can also explain matter-antimatter asymmetry as a consequence of gravitational anomalies, since the latter lead to undiluted KR backgrounds which violate CP-symmetry and lead to the leptogenesis/baryogenesis scenario.  Even though these microscopic details are not of our main concern here, they may play an important role for the completion of the RVM picture of the cosmological evolution.
We refer the reader to Ref.\,\cite{Anomaly2019a} for the technical details and specialized references, and to \cite{GRF2019} for a summary of the underlying framework. Here we will focus only on the thermodynamical aspects of the cosmic evolution with a vacuum energy density of the form (\ref{eq:rhoLambdaunified}) containing both the $H^4$ and $H^2$ dynamical components.
After inflation has taken place and $H\ll H_I$,  the $H^2$ term of the vacuum energy density (\ref{eq:rhoLambdaunified}) takes its turn and provides the main dynamical behavior of the vacuum energy density. In the next section we describe such behavior, which comprises also the evolution at the present time. Taken together, it suggests that the entire history of the universe can be described in an effective RVM language upon starting from the (bosonic part of the) effective action of string theory based on the massless gravitational multiplet. We, therefore, think that the RVM is worthy of a detailed discussion of its phenomenological implications, and in particular of its thermodynamical properties.

\subsection{RVM for the current universe}\label{RVMCurrentUniverse}
Obviously, the presence of the  ${\cal O}(H^4)$ terms in the vacuum energy density (\ref{eq:rhoLambdaunified})  can have an influence in the early universe (mainly during inflation), as it will be shown in the next section. However,  for the post-inflationary universe we can set $\alpha=0$, as the constant term $c_0$ and the dynamical component
 ${\cal O}(H^2)$   suffice.  Therefore, for the analysis of the current observations one can take the simplified form
\begin{equation}\label{eq:rLH}
\rL(H)=\frac{3}{\kappa^2}\left(c_0 + \nu H^{2}\right)=\rLo+\frac{3\nu}{\kappa^2}\,(H^{2}-H_{0}^{2}),
\end{equation}
where we have normalized such that the quantity $\rL(H_0)=\rLo\equiv\CC/\kappa^2=(3/\kappa^2)(c_0+\nu H_0^2)$ is just the current value of the vacuum energy density.  The coefficient $c_0$ is therefore related to the ordinary cosmological parameters as follows:
\begin{equation}\label{eq:c0}
  c_0=H_0^2\,\left(\OLo-\nu\right)\,,
\end{equation}
where $\OLo=\rLo/\rco$, with $\rco=3H_0^2/(8\pi G)$ the present critical density. The above formula is the canonical implementation of the RVM for the present universe and it shows that it predicts a certain degree of dynamics for the vacuum energy density provided $|\nu|$ is small but not negligible, what is corroborated by the fitting values to the latest observations, where one finds
$\nu\sim 10^{-3}$ -- see e.g. \,\,\cite{Sola:2016ecz,Sola:2017jbl,Rezaei2019}.

We can perform now the following thermodynamic considerations concerning the vacuum energy density evolution in the RVM around our time  The non-relativistic matter part of  (\ref{energyconservation2}) can be written as a differential equation in the scale factor as follows\,\footnote{As always, prime  indicates differentiation with respect to the scale factor, and we use the chain rule  $d/dt=a H d/da$}:
\begin{equation}\label{eq:conservationlaw3}
\drm(a)+\frac{3}{a}(1+w)\,\rmm(a)=-\dprL\,.
\end{equation}
We can set $w\to w_m=0$ in it since we treat non-relativistic matter strictly as dust.  From the vacuum energy density (\ref{eq:rLH}) we have $\dprL(a)=(3\nu/\kappa^2) dH^2/da $, and so using Friedmann's equation (\ref{eq:FriedmannEq}) we obtain $\rho_\CC'(a)=\frac{\nu}{1-\nu}\,\rho_m'(a)$. This allows us to rewrite the conservation law \eqref{eq:conservationlaw3} as a differential equation involving only the matter density:
\begin{equation}\label{eq:ODErhom}
\drm(a)+ 3\frac{1-\nu}{a}\rmm(a)=0\,.
\end{equation}
Its solution reads
\begin{equation}\label{eq:rhoma}
  \rmm(a)=\rmo\,a^{-3(1-\nu)}\,.
\end{equation}
Similarly, when radiation is dominant we replace $\rho_m\to\rho_r$ and   $w\to w_r=1/3$ in  Eq.\,\eqref{eq:conservationlaw3}, and the solution reads
\begin{equation}\label{eq:rhorad}
  \rr(a)=\rro\,a^{-4(1-\nu)}\,.
\end{equation}
Quite obviously, for $\nu\to 0$ the above conservation equations reduce to the standard ones.
Let us now consider the evolution of the vacuum energy density for nonrelativistic matter.
Upon inserting \eqref{eq:rhoma}  in  (\ref{eq:conservationlaw3}) and integrating  we find explicitly the vacuum energy density as a function of the scale factor in the matter-dominated epoch:
\begin{eqnarray}
\rho_\CC(a) &=& \rLo + \frac{\nu\,\rho_{m0}}{1-\nu}\left(a^{-3(1-\nu)}-1\right)\,. \label{eq:rhoVRVM}
\end{eqnarray}
We may now compute the interaction source $Q$ between matter and vacuum from equations (\ref{eq:rhoma}) and (\ref{eq:rhoVRVM}) in the matter-dominated epoch. Using any of these expressions in Eq.\,(\ref{eq:Qequations}), we consistently find
\begin{equation}\label{eq:QsourceMDE}
Q=3\nu H\,\rmm\,.
\end{equation}
We see that the RVM solution corresponds to an specific interaction source between matter and vacuum of the kind (\ref{Sources}), in contrast to other DVM's where it is assumed merely in an ad hoc manner.
 In other words, for the RVM the form of $Q$  appears automatically from the structure of the vacuum energy density evolution in that model or that of the corresponding  matter energy density.
The interaction source $Q$ being proportional to $\nu$  is responsible for the tiny anomaly ($|\nu|\ll1$) in the matter conservation law (\ref{eq:rhoma}) and at the same time allows the vacuum energy density  \eqref{eq:rhoVRVM}  to  evolve with the expansion.  From the previous energy density formulas and Friedmann's equation (\ref{eq:FriedmannEq}) we find the corresponding Hubble function in the current epoch:
\begin{equation}\label{eq:H2RVMlate}
  H^2(a) =H_0^2\left[ 1 + \frac{\Omega_{m 0}}{1-\nu}\left(a^{-3(1-\nu)}-1\right) \right]=\frac{H_0^2}{1-\nu}\left[ \Omo a^{-3(1-\nu)}+\OLo-\nu\right]\,,
\end{equation}
where $\Omega_{m 0}=\rmo/\rco$.  For $\nu=0$ we recover once more the standard Hubble function of the  $\CC$CDM. Notice that the sum rule for spatially  flat geometry reads as in the standard case: $\Omo+\OLo=1$, i.e. it does not depend on $\nu$.

\subsection{RVM for the early universe}\label{RVMearlyUniverse}

We shall now review the situation for the early universe.  In those times the dominant components of the universe were mainly relativistic (photons, neutrinos and any other particle that becomes relativistic at very high temperature). This feature is common to both the $\CC$CDM and the RVM. However, in contrast to the former, for the latter it is the vacuum state which dominates at the very beginning.  In fact, for the RVM there is no matter or radiation at the initial state, just vacuum energy. The huge amount of vacuum energy existing at that time triggers inflation and then decays into (relativistic) matter, so the universe became quickly populated by relativistic particles.  We can confirm this picture by solving explicitly the cosmological equations\,\cite{Lima:2012mu}. We start from Eqs.~(\ref{eq:FriedmannEq}) and (\ref{eq:PressureEq}) written as follows:
\begin{eqnarray}\label{12.1}
3H^2=\kappa^2(\rho_r+\rho_\Lambda(H))
\end{eqnarray}
\begin{eqnarray}\label{12.2}
3H^2+2\dot H=\kappa^2(\rho_\Lambda(H)-\frac{1}{3}\rho_r).
\end{eqnarray}
Here $\rL(H)$ is  the vacuum energy density given in (\ref{eq:rhoLambdaunified}).  At this early stage of the universe, we can neglect the constant term $c_0$ in that expression as it plays a role only in the late time universe, as shown in the previous section. In contrast, we must keep now the $\sim H^4$ term, which is irrelevant in the late epochs but becomes the chief contribution early on.
In order to get the expressions for  $\rho_r(a)$ and $\rho_\Lambda(a)$ as functions of $a$, we can solve for $H(a)$ in advance. Substituting (\ref{eq:rhoLambdaunified}) with $c_0=0$ in Eqs.~(\ref{12.1}) and (\ref{12.2}), we arrive at a differential equation for $H(a)$:
\begin{eqnarray}\label{14}
\dot H+2H^2=2(\nu H^2+\alpha\frac{H^4}{H_I^2})\,.
\end{eqnarray}
Using $\dot H=a H H'$,  the solution of the Hubble rate in terms of $a$ is easily found to be
\begin{eqnarray}\label{15}
H(a)=\sqrt\frac{1-\nu}{\alpha}\,\frac{H_I}{\sqrt{1 +D  a^{4(1-\nu)}}}\,,
\end{eqnarray}
where $D>0$ is a positive constant since $H$ must decrease with the expansion.  The sign of $\nu$ is unimportant at this stage, but as we shall see we must have $|\nu|\ll1$.  The initial value of the Hubble rate, $H(0)=\sqrt\frac{1-\nu}{\alpha}{H_I}\simeq {H_I}/\sqrt{\alpha}$, is larger than $H_I$. Most important, it is finite and hence there is no singular initial point. To insure this, it is indispensable that $\alpha>0$. Notice, however, that the limit $\alpha\to0$ of the solution \eqref{15} is not defined. The reason is that in this limit no non-singular solution  of Eq.\,\eqref{14} can exist at $a=0$,  except the trivial one  ($H=0$), as can be easily checked. In other words, it is only when the term $H^4$ is present, and carrying a positive coefficient, that nonsingular solutions to that equation can exist. The generalization to $H^{n+2}\, (n\geq1)$ can be found in the Appendix B.

With the above found expression for $H(a)$, we find from ~(\ref{12.1}) and (\ref{12.2}) the explicit forms of $\rho_r$ and $\rho_\Lambda$ as a function of the scale factor:
\begin{eqnarray}\label{16.1}
\rho_r(a)&=&\frac{3 H_I^2  (1-\nu)^2 D a^{4(1-\nu)}}{\kappa^2 \alpha (1+ D a^{4(1-\nu)})^2}
\end{eqnarray}
and
\begin{eqnarray}\label{16.2}
\rho_\Lambda(a)&=&\frac{3H_I^2(1-\nu) \left(1+\nu D a^{4(1-\nu)} \right)}{\kappa^2 \alpha \left(1+D a^{4(1-\nu)}\right)^2}.
\end{eqnarray}
Once more we see that the condition $\alpha>0$ is essential for the meaningfulness of these expressions. The bulk of the decaying process of vacuum into radiation is indeed controlled by the $H^4$-term (with coefficient $\alpha$), whereas the effect of $H^2$ (whose coefficent is $\nu$) is subleading at this stage\footnote{In fact, we could set $\nu=0$ to describe most features of the early universe, but as we shall see for some others it is convenient to track its exact dependence,  so we shall keep the $\nu$-dependence only throughout  the main formulas unless stated otherwise.}.
Let us now verify the announced results. First, we note that the above energy densities are well defined at $a=0$, which corroborates that there is no initial singularity.  Second, we confirm that $\rho_r(a=0)=0$ and $\rho_\Lambda(0)\simeq 3H_I^2/(\kappa^2 \alpha)>\rho_\Lambda(a)$ for all  $a>0$ (i.e. we have no radiation energy in the beginning while we have maximum, finite, vacuum energy density at that point.  Subsequently, for  small values of $a$ (i.e. in the very early stages),  the radiation density increases fast (specifically as $\sim a^{4(1-\nu)}$). The increase in the radiation density is brought about by the corresponding decay of the vacuum  into radiation, but the density of the vacuum state is huge to start with and remains essentially constant in the beginning, which is responsible for a short period of very fast inflation. Third, upon further  increase of the scale factor we meet a  crucial  point $a=a_{\rm eq}$ where the two densities (of vacuum and radiation) equalize, $\rho_\Lambda(a_{\rm eq})=\rho_r(a_{\rm eq})$, and from here onward they both start decaying\,\footnote{The very early point $a_{\rm eq}$ should not be confused with the equality point between radiation and nonrelativistic matter, which we may call $a_{\rm EQ}$.  Of course, we must have $a_{\rm eq}\lll a_{\rm EQ}$, see later on.}. However, the radiation density remains always  dominant since for sufficiently large $a$ the two densities decay  as $\sim a^{-4(1-\nu)}$  but the vacuum density carries an additional factor of $\nu$. Thus, the post-inflationary vacuum remains always suppressed, $\left|\rho_\Lambda(a)/\rho_r(a)\right|\propto|\nu|\ll1$, and it cannot jeopardize at all any of the basic features characterizing the standard radiation-dominated epoch in the $\CC$CDM. Therefore, after the inflationary epoch has been terminated the running vacuum universe smoothly transits into the standard concordance picture of the cosmic evolution.

Note that the parameter $D$ in the above equations only appears in the coefficient of $a^{4(1-\nu)}$, so $D$ can be eliminated after rescaling $a$ at a relevant point.  It is natural to use the aforementioned equality point  $a_{\rm eq}$. Thus, according to this definition,  we can determine  $a_{\rm eq}$ in terms of the parameters $D$ and $\nu$, and by inverting such relation  we can eliminate $D$ in terms of $a_{\rm eq}$, with the result:
\begin{eqnarray}\label{aeq0}
D=\frac{1}{1-2\nu}\,a_{\rm eq}^{-4(1-\nu)}\equiv\astar^{-4(1-\nu)}\,,
\end{eqnarray}
where we have defined $\astar$; obviously $\astar$ is essentially equal to $a_{\rm eq}$ since $|\nu|\ll1$ but it is more convenient to use the former since the formulas simplify. Therefore we may call sometimes $\astar$ also the vacuum-radiation equality point.
Using the above relation in the Hubble function (\ref{15}) we can reexpress it in a more compact form as follows:
\begin{eqnarray}\label{hubbleeq0}
H(\ha)=\frac{\tHI}{\sqrt{1 +\ha^{4(1-\nu)}}}\,,
\end{eqnarray}
where we have defined the rescaled scale factor
\begin{equation}\label{eq:ahat}
  \ha\equiv \frac{a}{\astar}\,,
\end{equation}
and for convenience we have also defined a rescaled $H_I$:
\begin{equation}\label{eq:tHI}
\tHI=\sqrt\frac{1-\nu}{\alpha}\,H_I\,.
\end{equation}
Similarly, we can rephrase  Eqs.~(\ref{16.1}) and (\ref{16.2}) in a more convenient way in terms of the rescaled scale factor $\ha$:
\begin{eqnarray}\label{eq:rhorfinal}
\rho_r(\ha)&=&\trI (1-\nu)\,\frac{\ha^{4(1-\nu)}}{\left[1+  \ha^{4(1-\nu)}\right]^2}
\end{eqnarray}
and
\begin{eqnarray}\label{eq:rhoLfinal}
\rho_\Lambda(\ha)&=&\trI\, \frac{1+\nu \ha^{4(1-\nu)}}{\left[1+  \ha^{4(1-\nu)}\right]^2}\,.
\end{eqnarray}
In conjunction with \eqref{eq:tHI} we have defined $\trI$, which is related to $\tHI$  through the Friedmann's-like relation:
\begin{equation}\label{eq:rhoI}
\trI=\frac{3}{\kappa^2}\,\tHI^2\,.
\end{equation}
As we can see from  (\ref{eq:rhoLfinal}), the value of $\trI$ is nothing but the vacuum energy density at $a=0$:  $\rL(0)=\trI$, hence at the beginning of the inflationary epoch.

We can derive the numerical order of magnitude  of $a_{\rm eq}$ $\simeq \astar$ as follows.  In the asymptotic limit $\ha\gg1$ (i.e. $a\gg a_{\rm eq}$), the radiation density (\ref{eq:rhorfinal}) behaves as
\begin{eqnarray}\label{eq:rhorfinal2}
\rho_r(a)&=&\trI (1-\nu) \ha^{-4(1-\nu)}=\trI (1-\nu) \astar^{4(1-\nu)}\,a^{-4(1-\nu)}\,.
\end{eqnarray}
This is the usual behavior of radiation, $\rho_r(a)\sim\rho_{r0} a^{-4(1-\nu)}$, up to a tiny correction in $\nu$, which we already found in Eq.\,(\ref{eq:rhorad}).
So both equations must be the same because both must reproduce the same radiation density at present ($a=1$), which we denoted $\rho_{r0}$.  The current radiation contribution to the total energy budget is known,  $\Oro=\rro/\rco\sim 10^{-4}$, where $\rco\sim 10^{-47}$ GeV$^4$ is the critical density in natural units.  On the other hand the energy density at the inflationary period must be of order of the GUT one, $\trI\sim M_X^4$, with $M_X\sim 10^{16}$ GeV the typical GUT scale. Therefore, we find (neglecting the numerical effect of $|\nu|\ll1$):
\begin{equation}\label{eq:astar}
a_{\rm eq}\simeq\astar=
\left(\Oro\,\frac{\rco}{\trI}\right)^{\frac{1}{4(1-\nu)}}\simeq \left(10^{-4}\,\frac{10^{-47}}{10^{64}}\right)^{1/4}\sim 10^{-29}\,.
\end{equation}
Needless to say we cannot provide an exact numerical value of $\astar$  since it is model-dependent (e.g. it depends on the exact value of $\trI$ associated to a given GUT) and moreover one cannot reach the vacuum-radiation equality point assuming that the evolution is always within the radiation epoch; nonetheless the procedure provides an order of magnitude estimate~\cite{Sola:2015rra}. Recall that the equality point between radiation and nonrelativistic matter is much later, $a_{\rm EQ}\sim 3\times 10^{-4}$, so indeed we have found $a_{\rm eq}\lll a_{\rm EQ}$, as expected. Later on in our discussions about the entropy of the early universe (cf. Sec.~\ref{ProductionEarlyUniverse}), we will see the convenience of introducing such vacuum-radiation equality point.

\mysection{Particle and entropy production in comoving volume: current universe}\label{ProductionCurrentUniverse}

In the following we study particle and entropy production in the presence of dynamical vacuum. We illustrate it first using the RVM studied in Sec.~\ref{RVM}  for the current universe.  In this case the matter particles are nonrelativistic, say with mass $m$, and its number density is related to the energy density through $\rmm=m\,n$. The anomalous matter conservation law (\ref{eq:rhoma}) can be interpreted in different ways.  One possibility is that $n$ does not precisely follow the normal dilution law with the expansion, i.e. $n\sim a^{-3}$, but the anomalous law:
\begin{equation}\label{eq:nonconservednumber}
n(a)=n_0\,a^{-3(1-\nu)}\ \ \ \ ({\rm at\ fixed\ particle\ mass}\ \ m=m_0)\, \ \ \ ({\rm Scenario\ A})\,.
\end{equation}
But another possibility is to interpret that the particle masses $m$ do not stay constant with time and hence that they mildly
scale with the cosmic evolution:
\begin{equation}\label{eq:nonconservedmp}
m(a)=m_0\,a^{3\nu}\ \ \ \ ({\rm with\ normal\ dilution}\ \ n(a)=n_0\,a^{-3})\, \ \ \ ({\rm Scenario\ B})\,.
\end{equation}
Either one of these options can apply to all particles or to some particular species, for example to dark matter (DM) particles, which are the most abundant and therefore the impact can be largest. Here, we essentially focus on Scenario A since we mainly study the entropy production through the change in the number of particles from vacuum decay.  In fact,
during the expansion a certain number of particles  are lost into the vacuum, if $\nu<0$; or ejected from it (through vacuum decay), if $\nu>0$. We wish to study which one of these possibilities is more favored by thermodynamics.

Within Scenario A we find the following.  First we formulate the balance equation (\ref{balanceLaw}) for the number density of particles in terms of the scale factor rather than the cosmic time:
\begin{equation}\label{eq:balance2}
  n'(a)+\frac{3}{a}\,n(a)=\frac{\Gamma}{H a} n(a)\,.
\end{equation}
Substituting (\ref{eq:nonconservednumber}) in (\ref{eq:balance2})
it is easy to find
\begin{equation}\label{eq:Gamma3nu}
  \Gamma=3\nu\, H\,,
\end{equation}
where $H$ is given by Eq.\,(\ref{eq:H2RVMlate}) for the current universe.
This means that the particle production rate is entirely determined by the index $\nu$ and the expansion rate $H$, which is not surprising  given the fact that such index controls the very evolution of the vacuum energy density (see Eq.\,(\ref{eq:rhoVRVM})). The latter implies, for $\nu>0$, that such density decreases with the expansion (as it decays into particles). It is consistent with Eq.\,(\ref{eq:Gamma3nu}), which is telling that the particle production rate is positive in such case. We can readily crosscheck  the result (\ref{eq:Gamma3nu}) taking  Eq.\,(\ref{relation1}) as a starting point. In the matter-dominated epoch ($w_m=0$) we may confirm from Eqs. (\ref{eq:rhoma}) and (\ref{eq:rhoVRVM}):
\begin{eqnarray}
\Gamma=\frac{-\drL(t)}{(1+w_m)\rho_m}=\frac{-\rho_\CC'(a)\,a H}{\rmm(a)}=\frac{-3(1-\nu)\,\nu\,\rho_{m0}\,a^{-3(1-\nu)}\,H}{(1-\nu)\,
\rho_{m0}a^{-3(1-\nu)}}=3\nu\,H\,.\label{relation1b}
\end{eqnarray}
Mind that (\ref{relation1}) holds for  $\dot{\sigma}=0$. Thus, applying now  Eq.\,(\ref{eq:sigmaconstant}) we find that the entropy rate evolves as $\dot{S}_m/S_m=3\nu H$ or equivalently $S_m'/S_m=3\nu/a$, where the subindex $m$ denotes entropy associated to nonrelativistic matter particles. Therefore, the entropy increases for $\nu>0$ and decreases for $\nu<0$, which is once more consistent with the sign of the particle production rate in (\ref{eq:Gamma3nu}).  Integrating, we find
\begin{equation}\label{eq:Sevolution}
  S_m(a)=S_{m0} a^{3\nu}\,,
\end{equation}
where $S_{m0}=n_0\sigma$ is the current ($a=1$) entropy of matter particles per comoving volume.  The derivatives of \eqref{eq:Sevolution} are simply
\begin{equation}\label{eq:Sprimes}
  S'_m(a)=3\nu S_{m0} a^{3\nu-1}\,,\  \ \ \ \ \ \ \ \ \  S''_m(a)=3\nu (3\nu-1) S_{m0} a^{3\nu-2}\,.
\end{equation}
Because $\dot{\sigma}=0$ is assumed here,  and owing to  $|\nu|\ll1$  it follows from \eqref{eq:Sprimes} that only for $\nu>0$ we can achieve both $S_m'\geq0$ (Second Law)  and $S_m''<0$ (LTE). These are the basic couple of thermodynamic conditions (\ref{eq:TwoConditions}) insuring that the system (in this case the universe) is driven to a final state close to thermodynamic equilibrium.  One can easily check that the previous results \eqref{relation1b}-\eqref{eq:Sprimes} hold good also for vacuum decaying into radiation in the immediate post-inflationary epoch, i.e. when  $\rho_r\propto a^{-4(1-\nu)}$, although for the current universe one mainly focus on  vacuum decay into  DM particles.  We conclude that in the context of running vacuum models, and for constant specific particle entropy, the only sign for $\nu$ which is compatible with the basic thermodynamic postulates is $\nu>0$. It warrants a situation of vacuum decay into particles with an associate increase of entropy until practically reaching thermodynamic equilibrium. The latter could be strictly achieved if we were  eventually having $S_m'=0$ and $S_m''<0$\,\cite{BookCallen1960}. Strictly speaking, however, it is never completely reached since  $S'_m(a)$ stays positive and decreases more and more, but it never vanishes. Equilibrium is nevertheless approached in an asymptotic way. Interestingly enough, the recent and successful confrontation of the RVM against the overall cosmological data show that these models are competitive with those from the $\CC$CDM;  and they render $\nu\sim 10^{-3}>0$ as the characteristic order of magnitude of the running vacuum parameter~\cite{Gomez-Valent:2014rxa,Sola:2015wwa,Sola:2016jky,Sola:2017znb,
Sola:2016hnq,Sola:2016ecz,Sola:2017jbl,Rezaei2019}.

We note that the result (\ref{eq:Sevolution}) can also be expected from the fact that the total entropy of the matter particles in the comoving volume $a^3$ must be equal to the total number of material particles in it times the specific entropy per particle,  $\sigma$, which we assume constant in this case. That is to say, $S_m=N\sigma=n a^3\sigma$.  Since $n$ is given by  (\ref{eq:nonconservednumber}) we find once more  $S_m=n_0\sigma a^{3\nu}\equiv S_{m0} a^{3\nu}$. Alternatively, the same conclusion could be achieved if in the previous equation  we would associate the evolving factor $a^{3\nu}$ to $\sigma$ rather than to $n$, namely $\sigma=\sigma_0 a^{3\nu}$.  This last case could be considered a particular implementation of Scenario B  -- cf. Eq.\,\eqref{eq:nonconservedmp} -- in which the entropy evolution is ascribed to the particles themselves and not to their increasing number.

Finally, consider another application of Eq.\,\eqref{eq:Gamma3nu} above. If we combine it with Eq\,\eqref{eq:conservationlawMatter3} in the context of the  (nonrelativistic) matter-dominated epoch, where in the source function we have $\rho=\rho_m$, we find
\begin{eqnarray}\label{eq:Qdm}
\dot{\rho}+3H \rho=3\nu H\,\rho_m\,.
\end{eqnarray}
The above equation takes the form of the interaction source given in Eq.\,\eqref{eq:QsourceMDE}. Depending on the model, $\rho$ in the source function could just be the cold dark matter part, i.e. $\rho=\rho_{cdm}$.  However we still keep the full matter density $\rho$ on the \textit{l.h.s} of \eqref{eq:Qdm}. Its final form will depend on the assumptions that are made on the conservation of  baryons ($\rho_b$) and cold dark matter ($\rho_{cdm}$). If e.g. baryons are assumed to be conserved ($\dot{\rho}_b+3H \rho_b=0$), then $\rho$ on the  \textit{l.h.s}  of \eqref{eq:Qdm} would be $\rho_{cdm}$ (exactly as on the \textit{r.h.s}) i.e. just the cold DM component. However, we shall not adopt any of these particular scenarios, which do not affect our thermodynamical analysis. Retaking our line of thought with  Eq.\,\eqref{eq:Qdm}, we note that
if we now take both the contribution from non-relativistic ($\rho_m$) and relativistic ($\rho_r$)  matter on the \textit{r.h.s.} of \eqref{eq:conservationlawMatter3}  we find
\begin{eqnarray}
\dot{\rho}+3H \rho=\nu H\,\left(3\rho_m+4\rho_r\right)\,,\label{AnomalousConservLaw3}
\end{eqnarray}
where we have used $p_r=\rho_r/3$ for the relativistic component. This form of the interaction source adapts once more  to the general formula \eqref{Sources}, and it was considered e.g. in \cite{Sola:2017jbl}. The differences in the above scenarios can have an impact for the description of the cosmological observations, but do not affect the kind of thermodynamical considerations that we are analyzing, which bare relation with the long term evolution of the entropy in the universe.

\mysection{Particle and entropy production in comoving volume: early universe}\label{ProductionEarlyUniverse}
In this section, we shall deal with particle production and the corresponding entropy in the early universe.  Upon a straightforward calculation from Eq.~(\ref{relation1}) combined with  Eqs.~(\ref{eq:rhorfinal}) and (\ref{eq:rhoLfinal}) we find the expression for the particle production rate:
\begin{eqnarray}\label{17}
\Gamma(\hat a)=-\frac34\,\frac{\dot\rho_\CC}{\rho_r}=3 \tHI\, \frac{2-\nu+ \nu\, \hat a^{4(1-\nu)}}{\left[1+ \hat a^{4(1-\nu)}\right] \sqrt{1+\hat a^{4(1-\nu)}}}>0,
\end{eqnarray}
where we have used $w_r=1/3$ (as in this stage vacuum decays into relativistic particles) and the rescaled  quantities  \eqref{eq:ahat} and \eqref{eq:tHI}. Thus, we see that the particle production rate is positive and remains approximately constant  (as the vacuum energy density itself) in the initial stages  $\ha\ll1$. It acquires a very large value, $\Gamma\simeq 6 \tHI$,  since $H\simeq\tHI$ itself is very large in the early universe. For $\ha\gg 1$, however,  it behaves as
\begin{eqnarray}\label{17b}
\Gamma(\hat a\gg1)\simeq3\nu\, \frac{\tHI}{\sqrt{1+\hat a^{4(1-\nu)}}}=3\nu H,
\end{eqnarray}
where we have used (\ref{hubbleeq0}). Thus, in such limit the above expression takes on the form we had found previously, cf. Eq.\,\eqref{eq:Gamma3nu}, which was expected since such relation is presumably active in our epoch within the context of the  RVM.  As we can see, $\Gamma$ eventually  becomes much smaller than $\tHI$ and proportional to a rapidly decaying  expansion rate ($H\sim \tHI \ha^{-2}$)  times the coefficient $\nu$. As it turns out, the $\nu$ parameter, despite it being small, is  relevant in this case since it controls the asymptotic limit of the particle production rate, and for $\nu>0$ it insures that the  vacuum continues (mildly) decaying into particles even  in the post-inflationary era.

Let us now compute the radiation entropy of the produced relativist particles. We know that their energy density increases as the fourth power of the temperature, and their number density and entropy increase as the cubic power of the temperature.  More specifically,  the corresponding energy density  reads as follows:
\begin{eqnarray}\label{19}
\rho_r=\frac{\pi^2}{30}g_\ast T_r^4.
\end{eqnarray}
Here we have included the usual factor  $g_\ast$ counting the total number of effectively massless degrees of freedom\,\cite{KolbTurner:1990}. It follows that the radiation temperature is related with the radiation density as
\begin{eqnarray}\label{eq:Tr}
T_r=\left(\frac{30}{\pi^2 g_\ast}\right)^{1/4} \rho_r^{1/4} =\tTI\,(1-\nu)^{1/4} \frac{\ha^{(1-\nu)}}{\left[1+  \ha^{4(1-\nu)}\right]^{1/2}}\,.
\end{eqnarray}
\begin{figure*}
\includegraphics[width=8.3cm,height=7.3cm]{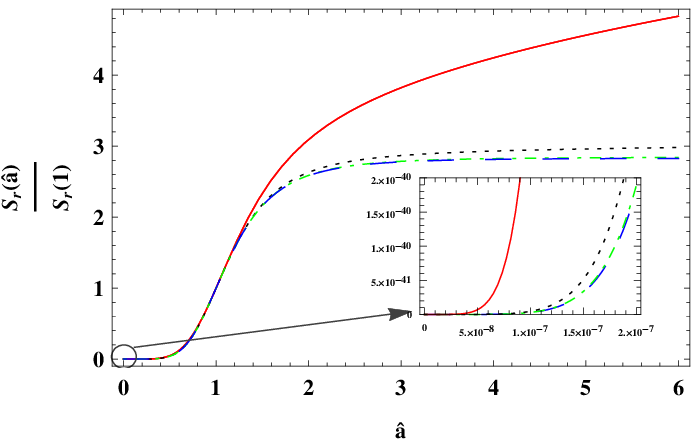}\\
\vspace{0.3cm}
\includegraphics[width=8.8cm,height=7.3cm]{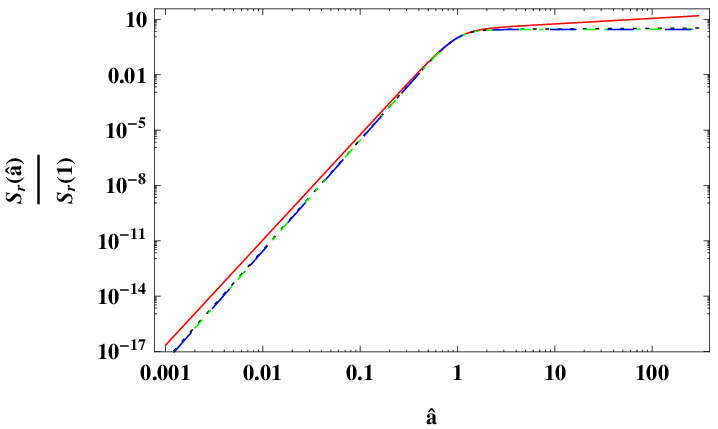}
\caption{\scriptsize Evolution of the radiation entropy in the comoving volume $V=a^3$ in the early universe, see Eq.~(\ref{eq:SrRVM}). Notice that the results have been normalized with respect to the value at $\ha=1$ (corresponding to vacuum-radiation equality, see the text).
Four different values of the parameter $\nu$ are considered in the plots: $\nu=0.1$ (red solid), $\nu=0.01$ (black dotted), $\nu=0.001$ (green dash-dotted), and $\nu=0.0001$ (blue dashed). The inner figure of the upper plot magnifies the region around $\ha=0$, where the raising of the curve is faster.  The lower plot is made in bilogarithmic scale and displays the saturation plateau, which is strict only for $\nu=0$. However, for $\nu\neq 0$ the saturation is not perfect owing to the asymptotic behavior \eqref{eq:fasymptot} of function (\ref{eq:fha}), which is not exactly a constant. However, among the four cases considered only for $\nu=0.1$ (the largest value) the departure becomes manifestly visible.  See more explanations in the text.}\label{draft2}
\end{figure*}
In the previous equation, we have associated a temperature $\tTI$  to the density $\trI$ following the same scheme as in \eqref{19}:
\begin{equation}\label{eq:tTI}
\trI=\frac{\pi^2}{30}\,g_\ast\,\tTI^4\,.
\end{equation}
This is merely a convenient definition. The  physical quantity is the vacuum energy density at the beginning of inflation, $\trI$.
The radiation entropy  in the comoving volume $V=a^3$ obtains from the standard expression  $S_r(\hat{a})=\frac{p_r+\rho_r}{T_r}a^3=\frac43\,\frac{\rho_r}{T_r}a^3$, where we assume that the evolution is around equilibrium and hence we do not consider the chemical potentials in Eq.\,(\ref{eq:entropydensity}).
With the help of Eqs~(\ref{eq:rhorfinal}) and \eqref{19} or \eqref{eq:Tr}, and the above definitions, we can express the radiation entropy for the RVM in the following alternative ways:
%
\begin{eqnarray}\label{eq:SrRVM}
S_r(\hat{a})=\frac{2\pi^2}{45}\,g_\ast T_r^3a^3&=&\frac43\left(\frac{\pi^2 g_\ast}{30}\right)^{1/4} \rho_r^{3/4} a^3=\frac{4\trI}{3\tTI}\,f(\ha)\,\astar^3=\frac{2\pi^2}{45}\,g_{*}\,\tTI^3\,\astar^3\,f(\ha)\,,
\end{eqnarray}
where we have defined the function
\begin{equation}\label{eq:fha}
  f(\ha)=(1-\nu)^{3/4}\,\frac{\ha^{6-3\nu}}{\Big[1+\ha^{4(1-\nu)}\Big]^{3/2}}\,.
\end{equation}
In  Fig.~1 we plot  the evolution of the radiation entropy \eqref{eq:SrRVM}  normalized to its value at the vacuum-radiation equality point \eqref{eq:ahat}, i.e. the ratio  $S_r(\hat{a})/S_r(\hat{a}=1)$.  We observe that the raising of the entropy is very fast in the beginning and then it saturates.  We can see from the behavior of the function \eqref{eq:fha}  that the fast raising of the entropy in the initial stages when $a\ll\astar$ (i.e. $\ha \ll1$) is as  $S_r\sim \ha^{6-3\nu}$.  But after we leave behind the equality point $\astar$  and enter  deep into the radiation epoch  (i.e. when $\ha \gg1$) the raise of entropy levels off and saturates to an asymptotic value:
\begin{equation}\label{eq:SrSaturation}
  S_r(\ha\gg1)\simeq \frac{2\pi^2}{45}\,g_{*}\,\tTI^3\astar^3(1-\nu)^{3/4}\ha^{3\nu}
  \equiv S_{r0} \ha^{3\nu}\,,
  \end{equation}
where we have used the asymptotic form of function (\ref{eq:fha}):
\begin{equation}\label{eq:fasymptot}
f(\ha\gg1)\simeq (1-\nu)^{3/4}\,\ha^{3\nu}\,,
\end{equation}
valid for large values of the scale factor well after the equality point between vacuum and radiation.  In the above limit, Eq.\,(\ref{eq:SrRVM}) tells us that $g_\ast T_r^3a^3=g_{*}\,\tTI^3 (1-\nu)^{3/4}\astar^{3(1-\nu)} a^{3\nu}$, or in other words, $g_\ast T_r^3a^3\propto a^{3\nu}$.
If $g_{*}$ does not change in the considered interval of cosmic evolution, we find
\begin{equation}\label{eq:Tanuconst}
T_r a^{1-\nu}={\rm const}.\ \ \ \ \ \ \ \ (a\gg\astar)\,.
\end{equation}
This result is similar to the usual scaling law of the radiation temperature up to a $\nu$-correction in the power of the scale factor. In practice, for sufficiently small values of $\nu$ we have the usual law $T_r a=$const and we can approximate \eqref{eq:SrSaturation} by a constant asymptotic limit as follows (except for very large values of $\ha$):
\begin{equation}\label{eq:saturation}
S_r(\ha \gg1)\rightarrow  S_{\infty}\equiv\frac{2\pi^2}{45}\,g_{*}\,\tTI^3\astar^3\,,
\end{equation}
and in the same limit
\begin{equation}\label{eq:saturationRatio}
\frac{S_r(\ha \gg1)}{S_r(\ha=1)}\rightarrow\frac{1}{f(1)}=2^{3/2}\simeq 2.83\,.
\end{equation}
This saturation plateau can be easily recognized  in Fig.~1 for the smallest values of $\nu$ considered there ($\nu=10^{-3}, 10^{-4}$). These are the more realistic ones. In fact, after comparing the RVM at low energies with the cosmological data, one finds a typical fit value of $\nu$ in the ballpark of $\nu\sim 10^{-3}$  \,\cite{Sola:2016ecz,Sola:2017jbl,Rezaei2019}. For $\nu=0.01$, and specially for $\nu=0.1$,  instead, the deviation from the saturation value can be significant. It can be seen perfectly well in  Fig.~1 (red curve).

But there is more to say concerning Fig.~1.  We can see by simple inspection that the radiation entropy generated in the RVM context  does indeed satisfy the Second Law of Thermodynamics  in any comoving volume.  This is confirmed  from the fact that the time- and scale factor-evolution of the entropy is a curve that starts convex ($S'>0$; $S''>0$) but becomes concave ($S'>0$; $S''<0$) around $\ha=1$. In other words,  it is a raising curve with a  slope initially increasing with the expansion ($S''>0$), but then at around the time of vacuum-radiation equality there is an inflexion point where the curve keeps on increasing but more moderately,  as  its slope starts decreasing  with the expansion ($S''<0$).  It follows that  beyond such inflexion point the two conditions for the fulfillment of the Second Law  ($S'>0$ and $S''<0$) are warranted, see  Sec.~\ref{sec2} and in particular Eq.\,\eqref{eq:TwoConditions}.

\begin{figure}
\centering
\includegraphics[width=11cm,height=7.5cm]{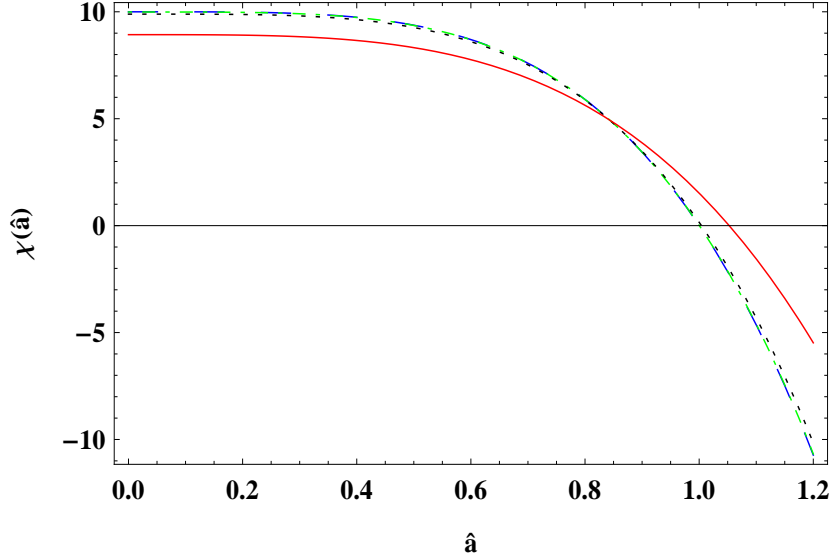}\\
\caption{\scriptsize Function  $\chi(\ha)$ defined in \eqref{eq:Sc}. As in Fig.~1. four cases of the parameter $\nu$ are presented in the plots: $\nu=0.1$ (red solid), $\nu=0.01$ (black dotted), $\nu=0.001$ (green dash-dotted), and $\nu=0.0001$ (blue dashed). The intersections of these lines with the horizontal coordinate are the inflexion points beyond which $S''<0$. When $\nu\ll1$, the value of $\nu$ has little effect on these points, which then lie at $\hat{a}\sim1$. This means that when $a\gtrsim\astar$ the second derivative of the radiation entropy starts to be negative. }\label{fig2}
\end{figure}

The behavior of the entropy curve in Fig.~1 is consistent with the Second Law and the LTE. Let us, however, verify it through an explicit analytical computation. The first and second derivatives of (\ref{eq:SrRVM}) can be taken with respect to the rescaled scale factor $\ha$.
As for the first derivative, we find
\begin{eqnarray}\label{21}
S_r'(\hat{a})&=&\frac{2\pi^2}{15}\,g_{*}\,\tTI^3\,\astar^3\,\frac{f(\ha)}{\ha}\,\frac{2-\nu+ \nu\, \hat a^{4(1-\nu)}}{1+ \hat a^{4(1-\nu)}}\,,
\end{eqnarray}
where $f(\ha)$ is the previously defined function \eqref{eq:fha}.
The second derivative can be written in the following compact form:
\begin{eqnarray}\label{22}
S_r''(\hat{a})&=&\frac{2\pi^2}{15}\,g_{*}\,\tTI^3\,\astar^3\,\frac{f(\ha)}{\ha^2}\,\frac{\chi(\ha)}{\Big[1+ \hat a^{4(1-\nu)}\Big]^2}\,,
\end{eqnarray}
where we have defined
\begin{equation}\label{eq:Sc}
\chi(\ha)=  (5-3 \nu)(2-\nu)-2(5-14\nu+7\nu^2)\hat{a}^{4 (1-\nu)}-\nu (1-3\nu)\hat{a}^{8 (1-\nu)}\,.
\end{equation}
It is not difficult to judge the sign of the first derivative of $S_r(\hat {a})$ from the expression \eqref{21} if we take into account that  $0<\nu\ll1$.  For any $\hat {a}$ we have $S_r'(\hat {a})>0$, which  meets  the first requirement of the Second Law of Thermodynamics.  On the other hand to judge the sign of   $S_r''( {\ha})$, note that in Eq.~(\ref{22}), all items, except $\chi(\ha)$, are positive definite. From (\ref{eq:Sc}) we easily see that the sign of $\chi(\ha)$ is positive, and approximately constant, for small $\ha$, but subsequently $\chi(\ha)$ decreases and eventually changes sign. Obviously, the inflexion point $\hat {a}_i$ can be exactly determined from the root of the function $\chi(\ha)$, i.e. $\chi(\hat {a}_i)=0$.  For  $|\nu|\ll1$ it is clear that $\hat {a}_i\simeq 1$  since the first and second terms on the \textit{r.h.s.} of \eqref{eq:Sc} almost cancel each other at $\ha\simeq 1$  while the third term is negligible  (again owing to  $|\nu|\ll1$).  The plot of  $\chi(\ha)$ is shown in Fig\,.\ref{fig2}.

From the above considerations we can draw the following conclusions: when $\hat {a}$ tends to zero, $S_r''(\hat {a})$ is positive (but finite), and as soon as $\hat {a}$ increases it can be anticipated that there exists an inflexion  point $\hat {a}_i$ near $1$ (whose precise value depends on $\nu$), beyond which $S_r''(\hat {a})$ becomes negative.

The above discussion is all about the very early universe. Usually, for this stage we only require the evolution of the radiation entropy to satisfy the first condition associated to the Second Law  ($S_r'(\hat {a})>0$), and we may need not impose the second  condition ($S_r''(\hat {a})<0$), which is the one that anticipates an eventual state of thermodynamic equilibrium. So for the early universe, even though $S_r''(\hat {a})$ is positive, we can still accept it as being reasonable, at least for a certain period. As we shall see, such temporary convexity property of the entropy function is actually necessary if the RVM is to solve the entropy problem of our universe. The rampant growth of radiation entropy in the beginning, which follows a high power of the scale factor, $S\sim \ha^6$  (we neglect here the small $\nu$-correction in such power), is sustained until the saturation value (\ref{eq:saturation}) is attained, i.e. until it reaches the plateau level in  Fig.~1. This fact insures that the accumulated amount of entropy per comoving volume generated from the rapid vacuum decay in the very early universe is transmitted without loss to our days, thus solving the entropy and horizon problems within the RVM. This is impossible, of course, in the $\CC$CDM (see Sec. \ref{VIID} for more details). Once such early period of frantic entropy production has passed, a consistent description can be obtained because  $S_r''(\hat {a})$ turns negative at some point.  The existence of such inflexion point insures that thermodynamical equilibrium will be eventually achieved. We have seen that such point always exists and it is close to the vacuum-radiation equality point $a_{\rm eq}\simeq\astar$ (see Fig.~\ref{fig2}). In the next section, we reconsider the growth of entropy when we replace the comoving volume with the apparent horizon, by considering both its interior and its surface. It is believed that the use of the apparent horizon rather than the comoving volume is more natural for the physical considerations, but it requires a generalization of the Second Law of Thermodynamics and hence a conceptual change that must be ultimately checked observationally. As we shall see, the theoretical differences that appear in this new thermodynamical approach are remarkable.

\mysection{Vacuum dynamics and the Generalized Second Law}\label{generalizedthermodynamiclaw}

\subsection{Cosmic horizons and entropy}\label{sec:CosmicHorizons}
In the previous sections, we have dealt with the entropy production in the early and late universe by considering the comoving volume $V=a^3$. However, the modern thermodynamic formulation of the expanding universe usually involves the Generalized Second Law (GSL) of thermodynamics, in which one takes the horizon rather than the scale factor as the characteristic length, and then one considers the evolution of both the entropy of radiation and of the matter particles inside the  horizon (i.e. the volume entropy  $\SV$ inside it)  together with  the area contribution from the horizon itself (i.e. the horizon entropy $\SA$).  The GSL was first formulated for black holes by Bekenstein and Hawking\,\cite{Bekenstein,Hawking1976}  after they discovered  that black hole horizons have entropy and temperature associated with them. This paved the way for  a complete thermodynamical formulation of black holes.  Later on these ideas were extended to the case of cosmological horizons\,\cite{Gibbons:1977mu,PCDavis1987}.   On the other hand such considerations bare relation with the subsequent formulation of the holographic principle\,\cite{tHooft1993,Susskind1995}, in particular the possible deep connection between gravitation and thermodynamics\,\cite{Jacobson1995,Padmanabhan2005}.   See also \cite{Bousso2002} for a review.

However, what is meant by horizon here? The subject of cosmological horizons was confusing until it was clarified by W. Rindler in a classic paper on the subject, where he distinguished between event and particle horizons\,\cite{Rindler1956} --- see also his book \cite{RindlerBook2001}.  In fact, this  is a nontrivial issue, which we shall briefly summarize here but we  address, of course,  the reader to the aforesaid references for more details. In some holographic formulations the horizon is understood as particle horizon, namely the visible region of the universe for a comoving observer at a given time:
\begin{equation}
\label{eq:eventH}
 \ell_{\rm p}(a) = a \int_{t_i}^t\frac{dt}{a(t)}= a \int_0^{a}\frac{da'}{a'^2H(a')}\,,
 \end{equation}
 where $t_i$ is the initial instant of time where the scale factor vanishes:  $a(t_i)=0$; for example,  $t_i=0$, but we can also have $t_i=-\infty$  (de Sitter phase).
The notion of particle horizon is particularly useful when we are  dealing with causality issues (cf. Sec.\ref{VIID}). If the universe has always been in decelerated expansion (radiation and matter dominated epochs), it does have a particle horizon (i.e. $ \ell_{\rm p}$ is finite). If, however, the universe starts from an accelerated period, then it no longer has particle horizon (since $ \ell_{\rm p}=\infty$).

We can also consider the complementary concept of event horizon, which is the boundary of the spatial region to be seen by the comoving observer, namely
\begin{equation}
\label{eq:eventH}
 \ell_{\rm ev}(a) =  a \int_{t}^\infty\frac{dt}{a(t)}=a\int^{\infty}_{a}\frac{da'}{a'^2H(a')}\,.
 \end{equation}
This integral diverges for flat and for open FLRW universes with no cosmological constant, and hence there is no event horizon for them (all future events are eventually accessible). In contrast, the integral is finite for the $\CC$CDM. If $a\gg1$, then $\ell_{\rm ev}(a)\simeq\sqrt{3/\Lambda}\equiv 1/H_\CC$, where $H_\CC$ is the asymptotic value of $H$. Thus, for accelerated universes ($\CC>0$) there is an event horizon which prevents to see all future events, but in contrast these universes have no particle horizon (if the de Sitter phase is also in the early times) and hence they do not suffer from any horizon problem associated to causality (cf. Sec.\ref{VIID}).

The above two definitions of horizon are obviously different concepts, the former refers to our knowledge of events in the past whereas the latter refers to the events in our future.

Still a third definition is highly convenient. For the modern thermodynamic discussion of the cosmological expansion one actually adopts  the so-called cosmological apparent (or gravitational)  horizon, $\ell_h(t)$, which is also different from the previous two. For a review, see e.g. \,\cite{Bousso2002} and references therein.  It has become more suitable than the event horizon, as the latter requires one to know the entire future evolution and causal structure of spacetime.
The apparent horizon has also the property that the holography based on it obeys the First Law of thermodynamics, in contrast to the particle horizon\,\cite{BakRey2000}; and moreover  there is natural gravitational entropy associated with it, as shown by the original Bekenstein-Hawking definition for the case of black holes and its  generalization to cosmological horizons.  It turns out that the apparent horizon is the largest admissible holographic surface for these entropy considerations\,\cite{Bousso1999}. Thus, apparent horizons are generally considered more suited for thermodynamic discussions than event or particle horizons\,\cite{BakRey2000,Bousso1999,Bousso2002,Bousso2005,CaiKim2005,AkbarCai2006,CaiCao2007,Faraoni2011,
FaraoniBook2015,Melia2018}.

The apparent horizon  has a precise topological interpretation\,\cite{NovikovFrolov,BakRey2000,Poisson2004}, namely a two-dimensional surface for which the outgoing orthogonal null geodesics have zero divergence (see below).  Technically,  the apparent horizon is the surface acting as boundary of the so-called antitrapped spacetime region $r>{\ell_h}$ (viz. the region where the families of ingoing and outgoing null geodesics orthogonal to that surface are both divergent, i.e. with positive expansion parameter $\theta>0$). In contrast to the Schwarzschild and Kerr black holes, however, where the apparent horizon is a membrane separating causally connected events from those that are not  (as a permanent event horizon),  the universe has an apparent horizon which  is not static but time-dependent,  $\ell_h(t)$.

Specifically, for FLRW universes with nonvanishing spatial curvature $k$ the expression for the apparent horizon is  the time-evolving quantity\,\cite{BakRey2000}
\begin{equation}\label{eq:apparentH}
\ell_h(t)=\frac{1}{\sqrt{H^2(t)+k/a^2(t)}}\,.
\end{equation}
This expression can be easily derived by noting that the FLRW metric with spatial curvature $k$ can be written as
\begin{equation}
\label{eq:FLRWk}
ds^2= -dt^2 + a^2(t) \left( \frac{dr^2}{1-kr^2} + r^2
d\Omega^2\right),
\end{equation}
where $d\Omega^2$ is the line element of the unit $2$-sphere. For $k=0$ we recover the flat space case \eqref{eq:FLRWmetric}, but let us keep $k\neq0$ for the present considerations. The metric (\ref{eq:FLRWk}) can be put in the alternative way

\begin{equation}
\label{FLRWk2}
ds^2 = h_{ab}dx^adx^b +\tilde r^2 d\Omega^2\,,
\end{equation}
in which one defines $\tilde r = a(t) r$ and $x^0=t$, $x^1=r$, with $h_{ab}$ the
2-dimensional metric $h_{ab} = {\rm diag} (-1, a^2/(1-kr^2))$.
The apparent horizon, being a trapped surface with
vanishing expansion, is defined in these variables through the null condition $h^{ab} \partial_a \tilde r \partial _b \tilde r=0$. Writing out this expression in the metric \eqref{FLRWk2} it is immediate to obtain the solution of that null equation,  $\tilde r_h(t)$. The radius of the apparent horizon is then $\ell_h(t)=a(t)\,\tilde r_h(t)$. Its explicit form is just Eq.\,\eqref{eq:apparentH}, as can be easily checked.

Physically speaking, we can say that beyond the cosmological apparent horizon all null geodesics recede from the observer and no information can reach us.   For spatially  flat spacetime ($k=0$) the apparent horizon boils down to  $\ell_h(t)=d_H=1/H(t)$, i.e. the Hubble distance or  radius of the ``Hubble sphere'' (or ``Hubble horizon'').  The latter is not a formal new kind of horizon\,\cite{FaraoniBook2015}, but just the limiting situation of the apparent horizon for vanishing spatial curvature. Still, the edge of the Hubble sphere/horizon, $d_H$,  is the ultimate lightcone for the observer since in it the galaxies recede at the velocity of light;  no event lying outside the Hubble sphere can ever be observed\,\cite{Harrison1991,HarrisonBook}.   When the weak energy condition is satisfied, though, the apparent horizon coincides with the event horizon or lies inside it (i.e. in this case generally  the event horizon is beyond the apparent horizon)\,\cite{NovikovFrolov}.  In the important case of the $\CC$CDM (where such energy condition is satisfied) the presence of the cosmological constant term, $\CC$,  is such that  the apparent horizon becomes eventually an event horizon, which then remains constant forever: $\ell_h=1/H_\CC=\sqrt{3/\CC}$.  In such case, the apparent horizon agrees with the cosmological event horizon of the de Sitter space (this is always so for pure de Sitter spaces with $k=0$).  Furthermore, the apparent horizon has a very important attribute: it always exists, in contrast to the event horizon (as we have seen above). In view of the aforementioned properties, we adopt  the apparent horizon as the basic gravitational  horizon for our thermodynamic discussion and it will be henceforth referred to for short just as horizon, with no further qualifications.

A similar situation occurs for the cosmology in which the vacuum is dynamical. We take once more as a prototype the RVM with flat space geometry. From  Eq.\,(\ref{eq:H2RVMlate}) we find  that the horizon in the remote future ($a\to\infty$) is
\begin{equation}\label{eq:horizonRVM}
\ell_h^{\rm RVM}(\infty)=\frac{1}{H_0\sqrt{1-\frac{\Omo}{1-\nu}}}
=\frac{1}{H_0\sqrt{\frac{\OLo-\nu}{1-\nu}}}\,,
\end{equation}
where $\OLo=1-\Omo$ for spatially flat FLRW metric. Notice that for $\nu=0$ it reduces to the the $\CC$CDM value,  $\ell_h^{\CC {\rm CDM}}(\infty)={1}/\left(H_0 \sqrt{\OLo}\right)=1/\sqrt{\CC/3}$,  as expected.   As for the entropy of the apparent horizon, at any given instant of time of the cosmological evolution, we use the Bekenstein formula introduced in the aforementioned papers:
$\SA=k_B\,{\A}/(4\ell_{P}^2)$, where $k_B$  is the Boltzmann constant (the unit of entropy), ${\cal A}=4\pi\ell_h^2$ is the area of the horizon,  and $\ell_P^2=\hbar/(M_P^2 c^3)$ is  the Planck length squared.  Roughly speaking, the entropy contribution $ \SA$ estimates  the number of times the elementary ``Planck area'' $\ell_P^2$  can be fit in the horizon area.  Notice that $k_B=1$ in natural units, similarly to $\hbar=c=1$, and in these units the entropy is dimensionless.    Since we continue adopting these units, we have $\ell_{P}^2=1/M_{P}^2=G$ and
\begin{equation}\label{eq:Bekenstein}
  \SA=\frac{A}{4G}=\pi\ell_h^2 M_P^2=\pi\,\frac{M_P^2}{H^2}\,,
\end{equation}
where in the last step we stick once more to flat space geometry $k=0$, which is the situation with which we started and the one it will be hereafter maintained.

In the following, we focus on the cosmological evolution of the sum of the  volume entropy inside the horizon and the entropy upon it: $S=\SV+\SA$.   We wish to ascertain if the  RVM is consistent with the GSL. That is to say, we wish to check if in  the transit from the different epochs of the cosmological evolution the following  generalization of the conditions (\ref{eq:TwoConditions}) within the GSL  holds good  or not:
\begin{equation}\label{eq:GSL}
\SVp+\SAp\geq 0\,,\ \ \ \ \ \ \ \ \ \ \ \ \ \ \ \ \ \  \SVpp+\SApp< 0\,,
\end{equation}
where in practice the differentiations will be with respect to the scale factor. For the early universe we will use the rescaled scale factor $\ha$, Eq.\,\eqref{eq:ahat} (being the most convenient variable in this case).

\subsection{Current universe evolving into the final de Sitter phase}\label{sec:CurrentFinaldeSitter}

\begin{figure*}
\includegraphics[width=11cm,height=6.5cm]{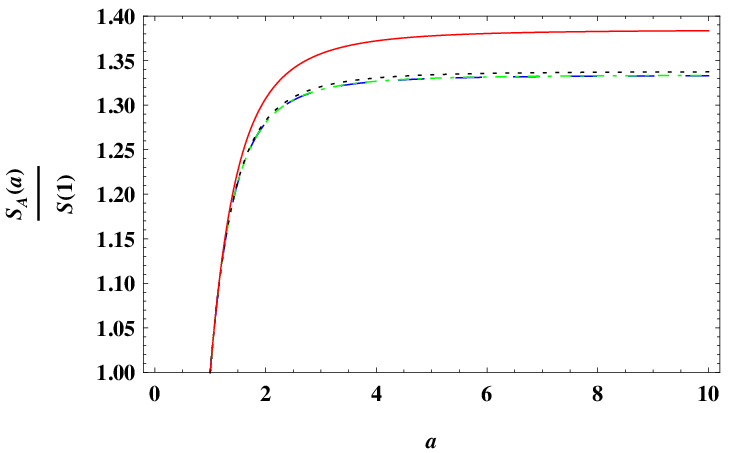}
\includegraphics[width=8cm,height=5cm]{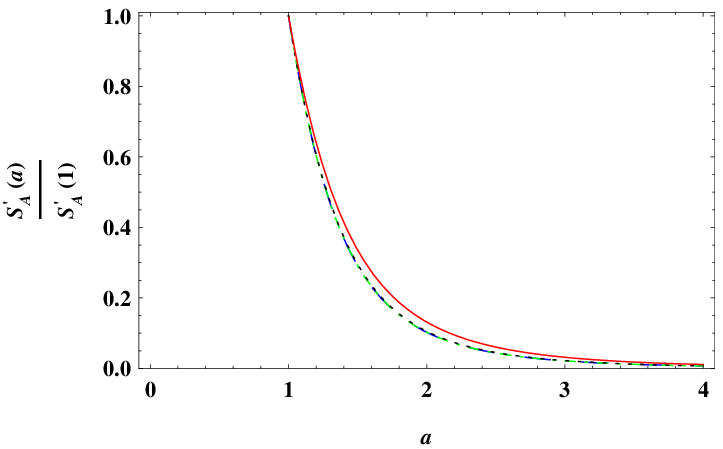}
\includegraphics[width=8cm,height=5cm]{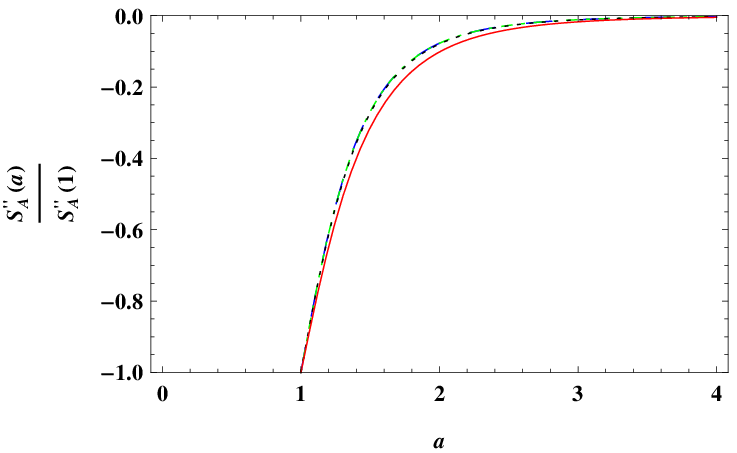}
\caption{\scriptsize Evolution of the the horizon entropy $\SA(a)$ (see Eq.~(\ref{7})) and its first and second derivatives, from the current universe into the future. The upper panel is the ratio $\SA(a)/\SA(a=1)$, normalized to its current value. The lower panels show the first and second derivatives ($S_{\cal A}'$, $S_{\cal A}''$) with respect to the scale factor, also normalized to the respective values at present. We can see that the two relations involved in the GSL, Eq.\,\eqref{eq:GSL}, are satisfied (recall that the numerical significance of $\SV$ is comparatively negligible, see the text). We plot these functions for four different values of the parameter $\nu$: $\nu=0.1$ (red solid), $\nu=0.01$ (black dotted), $\nu=0.001$ (green dash-dotted), and $\nu=0.0001$ (blue dashed).}\label{draft33}
\end{figure*}

To compute the horizon entropy from Eq.~(\ref{eq:Bekenstein}), we just need to know the expression of the Hubble function in each epoch. For the current universe, the Hubble rate of the RVM is given by Eq.~(\ref{eq:H2RVMlate}). Therefore the horizon entropy is
\begin{eqnarray}\label{7}
\SA(a)=\frac{\pi M_P^2}{H_0^2\left[1+\frac{\Omega_{m0}}{1-\nu}(a^{-3(1-\nu)}-1)\right]}.
\end{eqnarray}
As we shall see later, it is precisely because of the existence of the horizon entropy that in the late universe the total entropy in this context (the sum of the particle entropy and the horizon entropy) can satisfy the GSL (see Eq.~(\ref{eq:GSL})). The horizon entropy (\ref{7}) tends to a constant when the scale factor increases indefinitely: $\SA\to\frac{\pi M_P^2(1-\nu)}{H_0^2(1-\Omo-\nu)}$ (see Fig.~\ref{draft33}).

Although we have already calculated the particle entropy in the comoving volume (see Eq.~(\ref{eq:Sevolution})) and discussed its evolution, the particle entropy inside the apparent horizon is not a simple extension of Eq.~(\ref{eq:Sevolution}). When there is no particle creation and annihilation, the number of particles in the comoving volume is conserved, but the number of particles inside the horizon is not conserved. The volume entropy $\SV$  associated to nonrelativistic (dust) particles under the current horizon can naturally be defined in a similar way as for the comoving volume.  We know from Sec.\,\ref{ProductionCurrentUniverse} that in the latter case we have  $S_m=N\sigma=n \sigma V$, where $V=a^3$ is the comoving volume and $n$ the number density of particles.  In the present instance, in which we replace the comoving volume with the horizon volume, we define once more the entropy of the material  (nonrelativistic) particles contained in it as the product of their number density times the specific entropy per particle ($\sigma$) times the volume of the horizon ($V_h=(4\pi/3)\ell_h^3$):
\begin{equation}\label{eq:SV}
  \SV=n\sigma V_h=\frac{4\pi \sigma}{3}\,n\,\ell_h^3=\frac{4\pi \sigma}{3}\frac{n(a)}{H^3(a)}\,.
\end{equation}
In our case,  $n$ is a function of the scale factor given by  Eq.~(\ref{eq:nonconservednumber}). While for the comoving entropy we find that $S_m=n a^3\sigma=S_{m 0} a^{3\nu}$, where $S_{m 0}=n_0\sigma$ is the current comoving  value, the entropy of the material particles inside the horizon is quite different and  is controlled nontrivially by the Hubble rate rather than by the scale factor.  Notice also that whereas the comoving entropy stays constant for $\nu=0$  (as there is no vacuum decay into particles in such case), and remains approximately so for very small $\nu$,  the volume entropy inside the horizon does not stay constant at all, it depends on the Hubble rate and therefore evolves nontrivially.

Keeping in mind that we are assuming $\dot{\sigma}=0$ throughout our study, for simplicity we can just set $\sigma=k_B$ (Boltzmann constant, the unit of entropy), and hence in natural units $\sigma=1$. Then, taking Eqs.~(\ref{eq:H2RVMlate}) and (\ref{eq:nonconservednumber}) into the volume entropy formula above yields
\begin{eqnarray}\label{SV1}
\SV(a)=\frac{4 \pi  n_0  a^{-3 (1-\nu)}}{3}\left[H_0^2 \left(\frac{\Omega_{m0}}{1-\nu}(a^{-3(1-\nu)}-1)+1\right)\right]^{-3/2}.
\end{eqnarray}
The total entropy is given by the sum  $S_{total}(a)=\SV(a)+\SA(a)$, which is obtained from Eqs.~\eqref{7} and \eqref{SV1}.  Let us now analyze how the total entropy $S_{total}(a)$ evolves with the scale factor  when we consider the long time span from the current state of the universe into the final de Sitter phase. This is of course a very important period to analyze in order to assess if the long term thermodynamical evolution towards the future  is properly satisfied. The first and second derivatives of the total entropy, with respect to $a$, can be found after straightforward calculations:
\begin{eqnarray}\label{9}
S_{total}'(a)=\frac{(1-\nu)^2\pi a^{3\nu-4}}
{H_0^2\left[ \Omega_{m0}  a^{3 (\nu-1)}+\Psi\right]^2}
\Bigg[3 M_P^2 \Omega_{m0}+\frac{2  n_0 \left(\Omega_{m0}  a^{3(\nu-1)}-2 \Psi\right)}{\sqrt{H_0^2 \left(\frac{\Omega_{m0}  \left(a^{3 (\nu-1)}-1\right)}{1-\nu}+1\right)}}\Bigg]
\end{eqnarray}
and
\begin{eqnarray}\label{10}
S_{total}''(a)&=&\frac{-(1-\nu)^2 \pi a^{3 \nu-5}}{H_0^2 \left[\Omega_{m0}  a^{3 (\nu-1)}+\Psi\right]^3}\Bigg[3 M_P^2 \Omega_{m0}  \left[(3 \nu-2) \Omega_{m0}  a^{3 (\nu-1)}+ (4-3 \nu) \Psi\right]\nonumber\\
&+&\frac{n_0  \left[(3 \nu-1) \Omega_{m0} ^2 a^{6 (\nu-1)}+2 (14-15 \nu) \Omega_{m0} \Psi a^{3 (\nu-1)}+4  (3 \nu-4) \Psi^2\right]}{\sqrt{H_0^2 \left(\frac{\Omega_{m0}  \left(a^{3\nu-3}-1\right)}{1-\nu}+1\right)}}\Bigg]\,,\phantom{XXXX},
\end{eqnarray}
where we have defined the parameter
\begin{equation}\label{eq:Psi}
  \Psi=1-\nu-\Omega_{m0}=\OLo-\nu=\frac{c_0}{H_0^2}\,,
\end{equation}
and in the last step we have used Eq.\,(\ref{eq:c0}).
Notice that $\Psi\simeq \Omega_{\CC 0}\simeq 0.7$ because of the smallness of $\nu$. The positiveness of $\Psi$  is thus secured and it is of relevance for the subsequent discussion.
In order to determine whether the total entropy satisfies (\ref{eq:GSL}), we should keep in mind the order of magnitude of the  parameters in the expression for $S_{total}(a)$: $0<\nu \ll 1,~\Omega_{m0}\sim0.3,~H_0\sim 
10^{-42}$\,GeV  (in natural units)
and $n_0>0$.

We first analyze  $S_{total}'(a)$. The prefactor is of course positive and decreasing asymptotically as $\sim a^{-4}$. But the sign of the sum of the terms inside the square brackets must be considered attentively. The first item in the square brackets is a positive constant, $3 M_P^2 \Omega_{m0}$. The second item becomes negative with the evolution since $ n_0 \left( \Omega_{m0}  a^{-3}-2 \Psi  \right)\rightarrow  -2 n_0 \Psi<0 $. Here we only care about whether the expression (\ref{9}) satisfies $S_{total}'(a)>0$ when $a$ tends to infinity (which corresponds to the very late universe).  It is not difficult to realize that the limit of the second term is $\frac{-4n_0\Psi}{H_0\sqrt{1-\frac{\Omega_{m0}}{1-\nu}}}$ as $a\rightarrow \infty$.  Thus we need to compare the two terms in the bracket in the asymptotic limit, i.e.  $3 M_P^2 \Omega_{m0}$ and $\frac{-4n_0\Psi}{H_0\sqrt{1-\frac{\Omega_{m0}}{1-\nu}}}$. From the mentioned order of magnitude of the parameters involved, it is obvious that the following terms are of order one:  $3 \Omega_{m0}\sim\frac{4\Psi}{\sqrt{1-\frac{\Omega_{m0}}{1-\nu}}}\sim \frac{4\OLo}{\sqrt{1-\Omega_{m0}}}\sim 4\sqrt{\OLo}={\cal O}(1)$. Therefore, in the end we only need to compare $M_P^2$ with  $\frac{n_0}{H_0}$. Using Friedmann's equation, we have $H^2\sim \kappa^2\rho$. If we apply it to the current time and observe that $\rho_0=n_0 m$ (where $m$ is the average particle mass), we get the following estimate of the mentioned ratio:
\begin{equation}\label{eq:noHo}
  \frac{n_0}{H_0}\sim\frac{\rho_0}{m H_0}\sim\frac{H_0}{ m \kappa^2}\sim \left(\frac{H_0 }{ m}\right)\,M_P^2\lll M_P^2\,,
\end{equation}
where in the last relation we have taken into account that the current Hubble parameter expressed in mass units,  $H_0\sim 10^{-42}$\,GeV,  is much smaller than any known mass in the universe, i.e. $H_0/{ m} \lll1$ (except perhaps the mass of a hypothetical quintessence particle), but in any case much smaller than the average particle mass needed to account for the current mass density, including of course dark matter particles.
Summing up, when we move forward into the asymptotic future  ($a\rightarrow\infty$)  the first term in the square brackets on the \textit{r.h.s.} of Eq.\,(\ref{10}), i.e. the positive constant term  $3 M_P^2 \Omega_{m0}$, remains as the dominant term over the negative one proportional to $\frac{n_0}{H_0}$ , and therefore we are guaranteed that  $S_{total}'(a\rightarrow\infty)>0$,  in accordance with the GSL.  Put another way, it is thanks to the contribution from the horizon entropy, $\SA(a)$, that the GSL can be finally satisfied. In its absence,  the volume entropy from matter particles, $\SV(a)$, proves unable to do the job and hence it would entail a decrease of the entropy with the evolution!

Physically the reason lies on the fact that the particle decay rate of the vacuum in the current and future epoch follows the law \eqref{eq:Gamma3nu} and hence it is suppressed both by the smallness of $\nu$ and the decrease of $H$ during the evolution. As a result, the slow production of new particles cannot catch up with its rapid dilution owing to the expansion.  Alternatively, we can see from Eq.\,\eqref{eq:SV} that when we evolve into the future the value of $H^3$ in the denominator tends to a calculable constant from \eqref{eq:H2RVMlate} whose value is smaller than $H_0$,  whereas the numerator decays fast enough as in \eqref{eq:nonconservednumber}. The decrease of $H$ in no way can compensate for the steady drop of the number density of particles. In stark contrast,  $\SA(a)$ rather than decreasing it increases moderately with the expansion and compensates for it.

Next we need to determine the sign of $S_{total}''(a)$, which is essential for the stabilization of the entropy growth. Quite obviously, the prefactor on the \textit{r.h.s.} of Eq.\,(\ref{10})  is always negative and decreases as $\sim a^{-5}$, but the expressions inside the square brackets are a bit cumbersome and cannot be directly judged on a simple inspection. We can use the previous method to ascertain the sign of $S_{total}''(a)$ when the scale factor $a$ tends to infinity. We first find out the dominated items when $a\to\infty$ and ignore the subordinated terms. Using once more that  $|\nu|\ll1$, the eventual behavior  of the term  $(3 \nu-2) \Omega_{m0}  a^{3 (\nu-1)}+(4-3 \nu) \Psi$ is simply  $(4-3 \nu) \Psi$, which is positive. And the asymptotic behavior  of $(3 \nu-1) \Omega_{m0} ^2 a^{6 (\nu-1)}+2 (14-15 \nu-) \Omega_{m0}  a^{3 (\nu-1)} \Psi+4  (3 \nu-4) \Psi^2$ is $4  (3\nu-4) \Psi^2\simeq -16 \Psi^2$, which is negative.  Next we note that the limiting behavior  of the first term in the big square brackets gives the expression $3 M_P^2 \Omega_{m0} (4-3 \nu) \Psi\sim 12 M_P^2 \Omega_{m0} \OLo$, which is positive; whereas the  second term in the big square brackets  tends to  $\frac{4n_0(3\nu-4) \Psi^2}{H_0\sqrt{1-\frac{\Omega_{m0}}{1-\nu}}}\sim -16 \frac{n_0}{H_0}\,\OLo^{3/2}$, where the minus sign should be noted.  The two terms are with opposite signs and therefore  in competition. However, similar to the previous analysis, it is immediate to convince oneself that when $a\to\infty$  the items that need to be compared once more are $M_P^2$ and $\frac{n_0}{H_0}$ since the remaining factors  are of the same order of magnitude. Therefore, from the previous analysis of (\ref{eq:noHo})  we know that the former is much larger than the latter, so the overall expression in the big square brackets on the \textit{r.h.s.} of Eq.\,(\ref{10}) remains positive when me move towards the future and hence  $S_{total}''(a\rightarrow\infty)<0$.  This negative sign confirms the goodness of the asymptotic behavior  of the second condition  for the GSL and hence secures the stabilization (gradual decrease) of the entropy production.

At the end of the day we have proven in quite some detail  that the two necessary conditions  $S_{total}'(a\rightarrow\infty)>0$ and $S_{total}''(a\rightarrow\infty)<0$ are fulfilled for the total  entropy evolution of the RVM universe from the current time into the future, i.e. Eqs.\,(\ref{eq:GSL}). The requirements of thermodynamics for the asymptotic evolution are therefore satisfied both for the entropy in a comoving volume treated in a previous section and the entropy contribution from the horizon and its inside.  The conclusion is in accordance with previous analyses of entropy production for comoving volume in dynamical vacuum models \,\cite{Lima:2012mu,Perico:2013mna,JSPRev2013,Sola:2015rra,GRF2015}. Let us mention  also~\cite{MimosoPavon2013}, where the GSL is  explored for some models, including  a brief treatment of the RVM; and \cite{EspinozaPavon2019}, where a preliminary attempt is made at comparing the GSL expectations with observations  by pure (model-independent) kinematical analysis, rendering positive conclusions.  Here we have provided a full fledged theoretical account of the thermodynamical calculations for the RVM universe both for the comoving entropy and for the entropy associated to matter particles and horizon in the different epochs of the cosmic evolution and in the light of the existing fitting results of the RVM confronted to the overall cosmological data. In what follows we analyze how the horizon entropy of the universe evolves in the early epoch.

\subsection{From early inflation to the radiation-dominated epoch}\label{sec:InflationRadiation}

We have analyzed this period of the cosmic evolution in Sec.~\ref{ProductionEarlyUniverse} insofar as concerns the comoving entropy. But here, similarly to the previous section, we wish to account for the entropy evolution in the early universe when we consider the total entropy inside the horizon added to the entropy on the horizon, i.e. from the point of view of the GSL.
From the early de Sitter stage to the radiation-dominated epoch, the horizon entropy of the universe follows from equations \eqref{hubbleeq0} and \eqref{eq:Bekenstein}:
\begin{eqnarray}\label{23}
\SA(\hat {a})=\pi\,\frac{M_P^2}{H^2}=\frac{\pi  M^2_P  \left[ 1+\hat{a}^{4(1-\nu)} \right]}{\tHI^2},
\end{eqnarray}
where we use the same definitions as before. The growing of the horizon entropy in the beginning is very simple, just approximately constant:
\begin{equation}\label{eq:SAGrowth}
  \SA(\ha\ll1)\simeq\pi\,\frac{M^2_P} {\tHI^2}={\rm const.}\,,
\end{equation}
whereas in the asymptotic regime into the radiation epoch ($1\ll\ha\lesssim a_{\rm EQ}/\astar$) we have
\begin{equation}\label{eq:SAGrowth2}
  \SA(\ha\gg1)\sim \ha^{4(1-\nu)}\,,
\end{equation}
and therefore increases  very fast.  This result is consistent with the fact that when we enter deep into the radiation phase  ($\ha\gg1$) the Hubble rate should follow also from  Friedmann's equation    $H^2(a)=\rr(a)/3\kappa^2$, with  $\rr(a)$  given by Eq\,\eqref{eq:rhorad}, and we can check it renders the same result \eqref{eq:SAGrowth2}.
At the same time there is the  contribution  from the radiation entropy inside the horizon, which we also call the volume entropy in this context.
It is obtained from an expression similar to that in \eqref{eq:SrRVM} but with the comoving volume replaced by the horizon volume $V_h=(4\pi/3)\ell_h^3=4\pi/(3 H^3)$. Using the formulas  \eqref{hubbleeq0} and \eqref{eq:rhorfinal}, we find
\begin{eqnarray}\label{110}
\SV(\hat {a})&=&\frac43\left(\frac{\pi^2 g_\ast}{30}\right)^{1/4} \frac{4\pi}{3}\frac{\rho_r^{3/4}}{H^3(a)}=\frac{8\pi^3}{135}\,g_{\ast}
\left(\frac{\tTI}{\tHI}\right)^3(1-\nu)^{3/4}\,\ha^{3(1-\nu)}\,.
\end{eqnarray}
\begin{figure*}
{\label{Fig4}
\includegraphics[width=7.4cm,height=5.4cm]{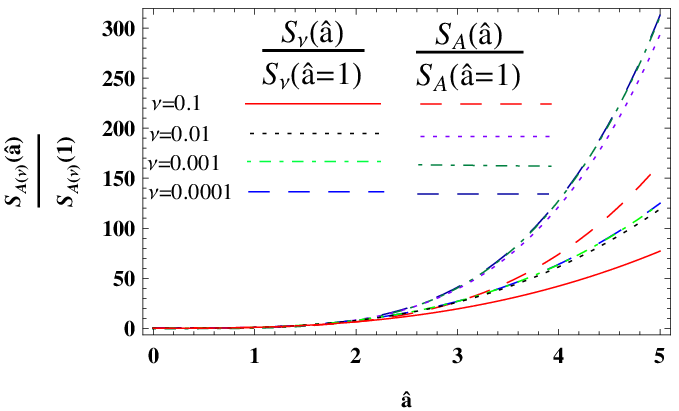}
\includegraphics[width=7.4cm,height=5.9cm]{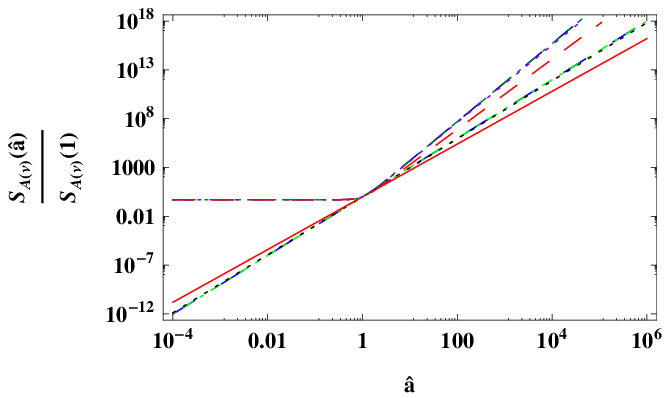}
\includegraphics[width=7.4cm,height=5.9cm]{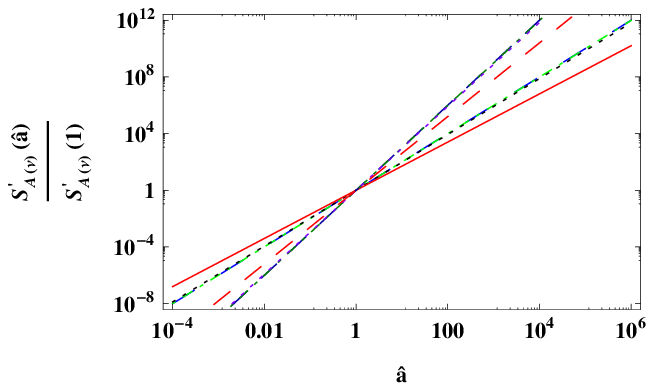}
\includegraphics[width=7.4cm,height=5.9cm]{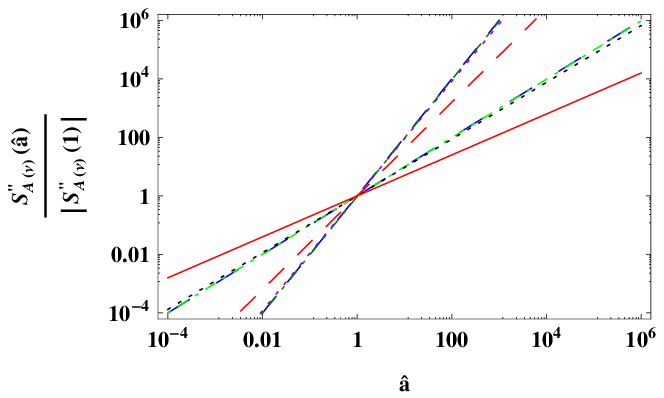}}
\caption{\scriptsize The evolution of the horizon entropy and radiation entropy (inside the horizon) from early inflation to the radiation-dominated epoch, see Eqs.~(\ref{23}) and (\ref{110}). The behavior of the radiation entropy here is very different from the radiation entropy in the comoving volume. While the raise of entropy in the last case wanes with time and tends to saturate (cf. Fig.~\ref{draft2}), the entropy on and inside the apparent horizon keeps growing in the early universe.
The two plots in the upper panel correspond to $\SV(\hat {a})/\SV(\hat {a}=1)$ and $\SA(\hat {a})/\SA(\hat {a}=1)$. The second plot is the same as  the first but in bilogarithmic scale, which helps to better observe the differences between the two contributions for $\hat {a}\to 0$. In particular, in the case of $\SA(\hat {a})$ one can spot an initial (approximate) plateau in $0\leq\ha<1$, whose value is defined roughly by \eqref{eq:SAGrowth}. The two plots in the lower panel are on the first and second derivatives of $\SV(\hat {a})/\SV(\hat {a}=1)$ and $\SA(\hat {a})/\SA(\hat {a}=1)$ with respect to $\hat a$, respectively. There is an interchange of slope dominance among the two contributions near $\ha=1$, which is connected with that initial plateau with very small slope. 
We plot four different values of the parameter $\nu$: $\nu=0.1$ (red solid and red dashed), $\nu=0.01$ (black dotted and purple dotted), $\nu=0.001$ (green dash-dotted and dark green dash-dotted), and $\nu=0.0001$ (blue dashed and dark blue dashed). In each set of parameters, the previous line stroke corresponds to the radiation entropy, and the later one is to horizon entropy.}
\end{figure*}
From the previous results, the total entropy inside the horizon, $S_{total}(\hat{a})=\SV(\hat{a})+\SA(\hat{a})$ is just
%
\begin{eqnarray}\label{24}
S_{total}(\hat{a})=\frac{\pi  M^2_P  \left[ 1+\hat{a}^{4(1-\nu)} \right]}{\tHI^2}
+\frac{8\pi^3}{135}\,g_{\ast}
\left(\frac{\tTI}{\tHI}\right)^3(1-\nu)^{3/4}\,\ha^{3(1-\nu)}\,.
\end{eqnarray}
The first and second derivatives read
\begin{eqnarray}\label{25}
S_{total}'(\hat{a})=\frac{4 \pi (1-\nu) \hat{a}^{2-3\nu}}{45\tHI^3} \left[2 \pi^2 g_{\ast} \tTI^3 (1-\nu)^{3/4} +45 \tHI M_P^2\hat{a}^{1-\nu}\right]
\end{eqnarray}
and
\begin{eqnarray}\label{252}
S_{total}''(\hat{a})=\frac{4 \pi  (1-\nu) \hat{a}^{1-3 \nu}}{45\tHI^3} \left[2 \pi ^2 g_{\ast} \tTI^3 (2-3 \nu) (1-\nu)^{3/4}+45 \tHI M_P^2(3-4 \nu)\hat{a}^{1-\nu}\right].\phantom{xxxxx}
\end{eqnarray}
In Fig.\,4 we plot the two contributions $\SV(\ha)$  and $\SA(\ha)$ from the early inflation epoch to the radiation-dominated epoch, as well as the the corresponding first and second derivatives with respect to the scale factor, i.e. $S_{\cal V,A}'$ and  $S_{\cal V,A}''$, all of them normalized with respect to the corresponding value at the equality point  $\ha=1$, similar to Fig.\,1.  It can be seen that the horizon entropy dominates ($\sim \ha^4$) over the radiation entropy inside the horizon ($\sim \ha^3$). Furthermore, since $0<\nu\ll 1$, it is easy to judge that the results of the derivatives taken together,  $S'_{total}(\hat {a})>0$ and $S''_{total}(\hat {a})>0$, imply an unrestricted growth of the entropy which violates the GSL, at least during the cosmological time when the above results are in force.

Such seeming violation of the GSL might be worrisome at first sight. Let us, however, stress that there is an important feature which still did not enter the above calculation since it is not relevant until  the late time universe, to wit:  the existence of a positive  cosmological constant, $\CC$ (connected with $c_0$ in Eq.~(\ref{eq:rhoLambdaunified})).  This will have a crucial impact in the sought-for thermodynamical equilibrium within the GSL, as we shall discuss in Sec.\,\ref{sec:Discussion}.

We close this section by comparing the results for the GSL  with the corresponding behavior that we have found for the comoving radiation entropy in Sec.~\ref{ProductionEarlyUniverse}, namely
\begin{equation}\label{eq:ExplosiveGrowth2}
  S_r(\ha\ll1)\sim\ha^{6-3\nu}\,, \ \ \ \ \ \ \ S_r(\hat{a}\gg1)=
   S_{r0} \hat{a}^{3\nu}\,,
\end{equation}
where the latter case reduces to Eq.~\eqref{eq:saturation} for $\nu=0$.  For $\nu\neq 0$, instead, it adopts the form discussed previously, Eq.\,\eqref{eq:Sevolution}. Since $|\nu|\ll1$ the evolution of the radiation entropy in the comoving volume is very small once we enter the asymptotic regime.  Such an evolution is nevertheless fully consistent with the Second Law provided $\nu>0$. Recall that equations (\ref{eq:Sevolution}) and (\ref{eq:Sprimes}) are also fulfilled for radiation.  In contrast,  for  the concordance $\CC$CDM the cosmic evolution in comoving volume  is isentropic, i.e. $  S_r(\ha)=$const. since for it  $\nu=0$.

Let us remark that the ``explosive'' early behavior of the asymptotic entropy in the context of the GSL framework ($\sim \ha^4$) is also present in the comoving volume case and it is even more pronounced ($\sim \ha^6$); it appeared as a convexity of the entropy function in the early period. Such behavior is actually useful for solving the entropy/horizon problems (see next section). But it was only a transitory feature, which  soon disappears owing to the presence of an inflection point, whereby the desired concavity behavior can rapidly be retrieved (cf. Fig\, 1).  Something similar will also happen in the GSL case, but it occurs only much later in the cosmic evolution.

\subsection{Solving the entropy/horizon problems in the RVM}\label{VIID}

Now that we have elaborated on the entropy evolution both in the comoving frame and from the point of view of the GSL upon making use of the notion of apparent horizon, let us discuss the famous entropy problem afflicting the $\CC$CDM\,\cite{KolbTurner:1990}, and its possible solution in the context of the RVM.   Let us  consider some point deep in the radiation epoch, fully in the adiabatic regime, and within the $\CC$CDM cosmology (hence $\nu=\alpha=0$).   In Sec.\,\ref{ProductionEarlyUniverse}, we have shown that for $\nu\to 0$ we  recover the entropy conservation law   $S_r=(2\pi^2/45)\,g_\ast T_r^3a^3=$const.  It therefore means that  this value must be equal  to the present value:
\begin{equation}\label{eq:S0}
S_{0}=
\frac{2\pi^2}{45}\,g_{s,0}\,T_{r0}^3\,\left(H_0^{-1}\right)^3\simeq
2.3 h^{-3} 10^{87}\sim 10^{88} \ \ \ \ \ \ (h\simeq 0.7)\,.
\end{equation}
The latter is the total (radiation) entropy enclosed in our physical horizon today. To compute it, we have replaced $a^3$ by the physical volume $a^3L^3$, which at present corresponds to $a_0^3 L_0^3$, with $a_0=1$ and  $L_0= H_0^{-1}$ being the current horizon; and we have used the known entropy factor    $g_{s,0}=2+6\times
(7/8)\left(T_{\nu,0}/T_{r0}\right)^3\simeq 3.91$
accounting for the  d.o.f. today, computed from the ratio of the present
neutrino and photon temperatures\,\cite{KolbTurner:1990}.  The large numerical value of the entropy \eqref{eq:S0} cannot be explained in the concordance $\CC$CDM model and leads to conflict when we retrace its origin\,\cite{KolbTurner:1990}.  To see it, we need to invoke the notion of particle horizon  (cf. Sec.\,\ref{sec:CosmicHorizons}), which is the one suitable for discussing the causality principle.  Using the $\CC$CDM cosmology to run the evolution back, starting from the  large entropy value \eqref{eq:S0}, let us consider any previous time $t$  where the scale factor is $a(t)$. Taking  $L=H^{-1}$ as a well known estimate of the particle horizon in the radiation and matter dominated epochs, we find
\begin{equation}\label{eq:St}
S(a)=
\frac{2\pi^2}{45}\,g_{\ast}\,T_r^3\,\left(H^{-1}(a)\right)^3\propto a^{-3}\left(H^{-1}(a)\right)^3\,.
\end{equation}
From  the standard expressions for $H$ in the $\CC$CDM model in the different standard epochs of the cosmic evolution, we  find that the entropy decreases dramatically inside the horizon when we move towards the past ($a\to 0$). The corresponding behavior in the relativistic and  nonrelativistic epochs are easily seen to be  $S\sim a^{3}$ and  $S\sim a^{3/2}$ respectively, so that  the number of causally disconnected regions when we move to the early  universe increases very fast. For instance, in the primordial nucleosynthesis epoch ($a\sim 10^{-9}$) the number of disconnected regions in the $\CC$CDM is $\sim 10^{27}$. This is of course a rephrasing of the horizon problem in standard cosmology\,\cite{KolbTurner:1990}.

In the context of the RVM, instead, the entropy and horizon problems can be resolved on account of the huge  entropy production in the early universe caused by the rapid  inflation and vacuum decay into relativistic matter. This is possible  thanks to the driving term $\sim H^4$ of the vacuum energy density of the RVM. Such term is active, in fact dominant, in the early universe if $\alpha\neq 0$ in Eq.\,(\ref{eq:rhoLambdaunified}).  Recall the rapid growth $S\sim \ha^{6}$ (the effect of $\nu$ can be neglected for this consideration) in the initial stages until it levels off and stagnates in a steady plateau as in  Fig.~1.  The accumulated amount of entropy per comoving volume generated from vacuum decay in the very early universe is eventually projected onto our days and can explain the huge number \eqref{eq:S0}. Recall that the large entropy generated at the end of the inflation period is transferred to the radiation phase, and then it is preserved by the standard evolution, up to a small $\nu$-correction -- see Eq.\,\eqref{eq:SrSaturation}. In this context, we can provide a consistent explanation of the large entropy today without clashing with the causal inconsistencies plaguing the $\CC$CDM description. Indeed, in the context of the RVM the particle horizon increases faster than the scale factor, it actually becomes infinite as we approach the very early stages of the cosmic evolution, in contrast to the  $\CC$CDM model. This can be easily checked as follows.  First note that for a pure de Sitter universe ($H=$const.), the particle horizon integral in Eq.\,\eqref{eq:eventH} does not converge for $a\to 0$. Such is in fact the case for the RVM near $a=0$, as can be seen from Eq.\,\eqref{hubbleeq0}. Thus, if we compute the ratio of the particle horizon and the scale factor we find
\begin{equation}
\label{eq:partHRVM}
 \frac{\ell_{\rm p}(a)}{a} = \int_0^{a}\frac{da'}{a'^2H(a')}=\lim_{\epsilon\to 0}\frac{1}{\tHI}\int_\epsilon^{a}\frac{da'}{a'^2}\sqrt{1 +\hat{a}'^{4(1-\nu)}}\to\infty\,.
 \end{equation}
The integration is performed over the range $0<\epsilon<a'<a$, and for finite $\epsilon$ the integral is well defined and positive in it. But it is easy to prove that it diverges for $\epsilon\to 0$.
This provides a solution of the horizon problem in the context of the RVM. In contrast, if we compute exactly the particle horizon  in the $\CC$CDM case we find that the corresponding ratio is finite and goes to zero for $a\to0$,
\begin{equation}\label{eq:partHLCDM}
\frac{\ell_{\rm p}(a)}{a}\propto  \int_0^{a}\frac{da'}{a'^2\left\{a'^{-2}, a'^{-3/2} \right\}}\to 0\,,
\end{equation}
both in the radiation ($H\sim a^{-2}$) and matter-dominated ($H\sim a^{-3/2}$) epochs. This is consistent with the drastic reduction of entropy in the $\CC$CDM  when moving backwards in time, as we have  estimated above.  Of course, this dramatic difference solves the horizon problem in the context of the RVM, as it implies that all the entropy that was generated in the very early stages of the cosmic evolution (cf. Sec. \ref{ProductionEarlyUniverse}) was produced under causal conditions before it was projected onto the future up to our days. In this way we can get a causal explanation for the big number \eqref{eq:S0}.

\mysection{The long way towards thermodynamical equilibrium}\label{sec:Discussion}

We have shown in Sec.~\ref{ProductionEarlyUniverse}  that when we consider the entropy evolution in the comoving volume for the early universe, there is a kind of `explosive growth' of entropy since the entropy  increases very fast in the beginning (roughly as  $S\propto\ha^6$).  It is actually welcome since we need such a huge generation of entropy during this stage in order to provide an explanation for the entropy problem of the $\CC$CDM in the context of the RVM, as we have explained in Sec.~\ref{VIID}.  In the comoving volume case, such overproduction of entropy is actually well under control already in the very early universe since it levels off rapidly after inflation, namely after attaining the vacuum-radiation equality point $\ha=1$ (cf. Fig.~1). The entropy continues increasing ($S'>0$), but less and less ($S''<0$). This  is fully  in accordance with the Second Law and the LTE.  However, if one adopts the notion of horizon and the GSL framework as  an starting point for the thermodynamical description,  we meet a different pattern of entropy growth which does not obviously conform with the thermodynamical expectations. Specifically, we find that  the surface contribution $\SA$  on the horizon provides from the start a huge amount of entropy whose growth  speeds up very fast, as $\SA\sim\ha^{4(1-\nu)}$, during the radiation dominated epoch. The radiation entropy inside the horizon (the volume entropy) also increases, but at a subdominant rate:  $\SV\propto\ha^{3(1-\nu)}$.  None of these rising trends actually satisfy the LTE since for them  $S_{_{\cal V},_{\cal A}}'(\ha)>0$ and  $S_{_{\cal V},_{\cal A}}''(\ha)>0$ (cf. Sec. \ref{sec:InflationRadiation}). They keep on that rhythm sustained  until reaching the matter-dominated epoch, where the  volume contribution inside the horizon suddenly drops down to  $\SV\sim a^{-3(1-\nu)}$,  corresponding  to the (non-relativistic) matter particles -- see Eq.~\eqref{SV1}.  On the other hand, the increasing rate of
the surface contribution $\SA$ is no longer  maintained in the form~\eqref{eq:SAGrowth2} at this point, but turns much smaller, as indicated by Eq.\,(\ref{7}). It is nonetheless still sufficient to lead the universe to safe thermodynamical equilibrium in the long run.

The path to satisfy the  LTE in the context of the GSL crucially depends on the  additive term  $c_0$ in Eq.\,(\ref{eq:rhoLambdaunified}) and hence on the existence of a constant contribution to the vacuum energy density, cf. Eq.\,\eqref{eq:c0}. For a clearer understanding, let us reexpress \eqref{7} as follows:
\begin{eqnarray}\label{eq:SAreason}
\SA(a)=\frac{\pi M_P^2\,(1-\nu)}
{H_0^2\,\left[\Omo\,a^{-3(1-\nu)}+\OLo-\nu\right]}.
\end{eqnarray}
If there would be no cosmological constant, or if we would be considering some instant in the cosmic evolution deep into the matter-dominated epoch ($a_{\rm EQ}\ll a\ll1$) prior to the dominance of the cosmological term, we would effectively have $\OLo-\nu=c_0/H_0^2\simeq 0$,  and in that case Eq.\,\eqref{eq:SAreason} would behave simply as
\begin{eqnarray}\label{eq:SAreason2}
\SA(a\ll1)\sim\,a^{3(1-\nu)}\,.
\end{eqnarray}
Such  behaviour of $\SA(a)$ tells us that during the matter-dominated epoch the path to thermodynamical equilibrium still does not conform to the GSL and looks divergent from it. So in the absence of a cosmological constant, the surface entropy of the horizon would be permanently increasing in the wrong, convex,  way ($\SA'(a)>0,\, \SA''(a)>0$).  Fortunately, there is a positive cosmological constant which emerges as a leading term in the late time universe, in particular at the present time, and the entropy formula \eqref{eq:SAreason2} is eventually replaced by \eqref{eq:SAreason}.
Around the time when this happens,  namely at the incipient dark energy epoch,  the horizon entropy sets off, finally, on the safe (concave) path towards thermodynamical equilibrium:  $\SA'(a)>0, \,\SA''(a)<0$ --  recall the first terms in the brackets of Eqs. \eqref{9} and \eqref{10},  which become dominant.   Once $\SA(a)$ behaves as in \eqref{eq:SAreason}, it  increases moderately up to a maximum asymptotic value $\SA(\infty)$ which is only  $(1-\nu)/(\OLo-\nu)\simeq 1.4$ times bigger than the value at present $\SA(a=1)$:
\begin{eqnarray}\label{eq:SAreason3}
\SA(\infty)=\frac{\pi M_P^2\,(1-\nu)} {H_0^2 (\OLo-\nu)}>\SA(1)=\frac{\pi M_P^2} {H_0^2}\,.
\end{eqnarray}
Such pace of moderate increase ($\SA'(a)>0$)  suffices to secure an evolution pathway with $\SA''(a)<0$  (i.e. the LTE) forever more, and thus conforms with the GSL  (cf. Fig. 3).   As it turns out, we find that it is the entropy contribution upon the horizon, $\SA$, the ultimate responsible for both the leading source of entropy during the early  inflationary phase as well as for guiding, finally, the universe into safe thermodynamical equilibrium at late times of its cosmological history. In the intervening period between the two de Sitter epochs of that history, the entropy of both radiation and of matter particles being stored inside the apparent horizon ($\SV$) did not play a decisive role in determining the universe's  fate as a macroscopic system subject to thermodynamical rules of evolution.

Summing up,  when we consider the entropy evolution from the point of view of the  GSL, the entropy on the horizon keeps on increasing from the very early de Sitter phase, it continues doing so in the radiation- and matter-dominated epochs and it only starts to moderate around the current and future eras when the presence of $\CC$ becomes dominant.  Thus, we conclude that it is the cosmological constant term, $\CC$,  the key actor which makes possible the safe way to thermodynamical equilibrium. It appears  as a rescue for the eventual fulfilment of the GSL in the late stages of the cosmic evolution, after the universe took enough time to recover from the  `explosive' growth of entropy acquired during the early de Sitter phase. But even after such a long path of cosmic evolution, the universe would never manage finding its way to safe equilibrium if no cosmological constant $\CC$ would exist, as the entropy would persist increasing in the untamed fashion (\ref{eq:SAreason2}) till the end of time.

All that said, let us emphasize that the peculiar behavior shown by the RVM when examined under the point of view of the  GSL is not so different from the concordance  $\CC$CDM model, since the latter is recovered from the former when there are no dynamical terms $H^2$ and $H^4$ in the vacuum energy density (\ref{eq:rhoLambdaunified}), that is to say,  when such energy density reduces to just a cosmological constant. Since those dynamical terms  in the vacuum energy cannot significantly affect the standard thermal history ($H^4$ decouples automatically after inflation,  and $H^2$ remains under control provided $|\nu|\ll1$), the  $\CC$CDM shares essentially the same sort of vicissitudes in connection to the GSL for the standard epochs of the cosmic evolution.  Notwithstanding, the RVM furnishes a radical change in the very early universe, as well as a small but measurable effect in the current epoch. First, it provides  a satisfactory  mechanism, based on $H^4$-driven inflation, which can solve the entropy/horizon problem of the $\CC$CDM (cf. Sec.\,\ref{VIID}); and second,  it leaves a mild $\sim\nu H^2$ dynamical imprint of the residual vacuum energy in the present universe, which may act  as a `smoking gun' for such possible completion of the cosmic history. Interestingly,  the form of dynamical dark energy predicted by the RVM  has been successfully tested in recent analyses of model observational data and points towards a better fit to the overall cosmological observations as compared to the $\CC$CDM\,~\cite{Sola:2015wwa,Sola:2016jky,Sola:2017znb,
Sola:2016hnq,Sola:2016ecz,Sola:2017jbl,Rezaei2019}.

\mysection{Conclusions}
\label{sec7}
In this paper, we have studied particle and  entropy production in the context of the running vacuum model (RVM), in which the vacuum energy density contains a constant terms and two dynamical contributions represented by the powers $H^2$ and $H^4$ of the Hubble rate.  Such canonical RVM form can describe the entire cosmic history successfully from the very early epoch of the universe (in which inflation is driven by the higher power $H^4$) until  the  current universe, when the vacuum energy density is essentially constant but it proves  still mildly dynamical owing to the lower power $H^2$.  The vacuum energy evolution is possible in these models since  the energy-momentum tensor of the matter is not conserved, which can be interpreted as production or annihilation of particles. Such dynamical interplay between the vacuum and matter is responsible for the nontrivial nature of the RVM as a generalized form of the $\CC$CDM in which the cosmological constant is replaced by a dynamical function of the Hubble rate which affects the entire cosmic history. We have remarked that these powers of the Hubble rate  can be the generic result of the low-energy effective action based on the bosonic gravitational multiplet of string theory, as shown recently\,\cite{Anomaly2019a,GRF2019}. The  presence of the  gravitational Chern-Simons term associated with that action turns out to lead to an effective $\sim H^4$ behavior when averaged over the inflationary spacetime. So $H^4$-driven inflation and the structure of the RVM can be the effective behavior of more fundamental theories. This is a good motivation to get deeper into its multifarious phenomenological consequences. In the Appendix B we have generalized our discussions of the entropy evolution in the context of a version  of the RVM beyond the canonical one, in which  inflation is driven by the higher order power of the Hubble rate  $H^{n+2}\, (n\geqslant 1)$ (preferably even powers $n=2,4,...$ owing to general covariance), and we have reached conclusions which are entirely similar to the case of $H^4$-driven inflation (corresponding to  $n=2$).

In the framework of the RVM  we have studied the particle entropy and radiation entropy evolution in the comoving volume approach during the different epochs, starting from the inflationary era. We found that the entropy satisfies the Second Law of Thermodynamics (i.e. $S'(\hat{a})>0$) and the LTE (law of Thermodynamic
Equilibrium:  $S''(\hat{a}\rightarrow \infty)<0$).  During the transition from the very early stage of the universe to the radiation-dominated period, the evolution of the entropy of ultra-relativistic particles is extremely fast in the comoving volume. Such `explosive' growth of radiation entropy only satisfies the Second Law, but not the LTE. This is actually welcome, to start with, since the large amount of produced entropy in that initial stage is very helpful to solve the entropy problem of the $\CC$CDM. Moreover, we have shown that such rampant growth of entropy  is automatically tempered thanks to  an inflexion point  around  vacuum-radiation equality (see Fig.~\ref{fig2}), beyond which the particle entropy  enters the radiation epoch and satisfies the LTE.  The existence of such point provides, of course, nothing but the  `graceful exit' mechanism in the context of the RVM description of the inflationary phase.

Finally, we have assessed the evolution of the entropy from the point of view of the Generalized Second Law (GSL) and the notion of apparent horizon. Here a new kind of problems have been  faced. The description turns out to be rather different as compared to the entropy in the comoving volume since the total  entropy contribution now,  $S_{total}(\hat{a})$, consists of two parts: the particle entropy within the horizon $\SV(\hat{a})$ and the horizon entropy $\SA(\hat{a})$. For the early universe ($\hat{a}\rightarrow0$), the total entropy  is mainly dominated by the entropy of the horizon surface  ($\SA(\ha)$) over the particle entropy inside the horizon ($\SV(\ha)$). When  the universe moves to the radiation-dominated epoch, it still carries an unrestrained entropy growth which does not show a hint of slowing down with the evolution.  As we have seen, the same sort of conclusions actually apply  to the $\CC$CDM.  Notwithstanding, for both the RVM and the $\CC$CDM, we found that the total entropy on the apparent horizon becomes fully in accordance with the Second Law as soon as we approach the current time and then into the future:  $S_{total}'(\hat{a}) > 0$ and $S_{total}''(\hat{a}\rightarrow\infty) < 0$.  It takes, however, the entire cosmic span prior to the current epoch until achieving  the long term purpose of  setting off towards thermodynamical equilibrium.  In contradistinction to the $\CC$CDM, the RVM provides a solution to the entropy problem, as it can provide a causal reason for the huge amount of entropy existing at present, Eq.\,\eqref{eq:S0}.

Remarkably, the existence of a nonvanishing, and positive, cosmological constant term is the crucial cosmic  ingredient directly responsible  for the eventual fulfilment of the GSL in the late universe and its subsequent evolution into the final de Sitter phase.  Thus, the cosmological constant term plays a key role here to salvage the cosmic evolution from failure to comply with the GSL mandate. One could say it is one important ``raison d'\^etre'' more for the cosmological constant in our universe.  As for its dynamical part at present, which evolves as $\sim \nu H^2$ within the RVM,  being $\nu>0$  it implies that the vacuum energy density is larger in the past than it is at present, and hence that  the vacuum throughout its cosmic evolution decays into  particles  (rather than the other way around). This is tantamount to saying that the dark energy that we see at present should be dynamical and behave quintessence-like, not phantom-like. Noteworthy, this is the kind of situation that is  hinted by the recent data fits to cosmological observations using the RVM\,~\cite{Sola:2016ecz,Sola:2017jbl,Rezaei2019} and also in generic parameterizations of the DE~\cite{PDU2019}.

Finally, we may ask ourselves which one of the two pictures should be adopted for a correct thermodynamical description of the evolution of our universe, that is to say,  the one based on the entropy associated to the comoving volume,  to which we apply the ordinary Second Law, or the one based on the apparent horizon, which is under the rule of the Generalized Second Law. This is a difficult question that we cannot fully answer at present, at least from first principles, since we still do not know if the universe as a whole can be treated as an ordinary macroscopic thermodynamical system.

\vspace{0.5cm}

{\bf  Acknowledgments}
\vspace{0.5cm}

\noindent
J.S.P is supported in part by  MINECO FPA2016-76005-C2-1-P, 2017-SGR-929 (Generalitat de Catalunya) and  MDM-2014-0369 (ICCUB). H.Y is partially supported by the National Natural Science Foundation of China (Grants Nos. 11875151, 11522541, and 11705070), and the Fundamental Research Funds for the Central Universities (Grants No. lzujbky-2018-k11), and also  by the scholarship granted by the Chinese Scholarship Council (CSC). He is grateful to the Departament de F\'\i sica Qu\`antica i Astrof\'\i sica and the Institute of Cosmos Sciences of the University of Barcelona for the warm  hospitality and support.

\newpage

\mysection{Appendix A: Alternative derivation of the entropy density relation (II.12)}
It is well-known\,\cite{BookCallen1960} that the entropy is a state function of $U,V$ and of all the particle number species, $N_i\,(i=1,2,...)$, what we indicate for short as $S=S(U,V,N_1,N_2,...)\equiv S(U,V,N)$.   Let us now rewrite  Eq.\, \eqref{eq:FirstLawThermo}  as
\begin{equation}\label{eq:FirstLawThermobis2}
dS=\frac{V d\rho+(\rho+p)dV-\sum_i\mu_i dN_i}{T}=\frac{\rho+p}{T}\,dV+\frac{V}{T}\frac{d\rho}{dT}\,dT-\sum_i\frac{\mu_i}{T}\, dN_i\,,
\end{equation}
where we used that $U$  is a function of the temperature at fixed $V$ and $N$, and hence the energy density is a function of $T$ at fixed $N$, $\rho=\rho(T)$.
 If we compare it with the full differential of the entropy,
\begin{equation}\label{eq:dSVTNi}
dS=\left.\frac{\partial S(V,T,N)}{\partial V}\right |_{T,N}dV+\left.\frac{\partial S(V,T,N)}{\partial T}\right |_{V,N}dT+\left.\sum_i\frac{\partial S(V,T,N)}{\partial N_i}\right |_{V,T}dN_i\,,
\end{equation}
we can derive obvious identifications for the values of the partial derivatives ${\partial S}/{\partial V}$, ${\partial S}/{\partial T}$ and ${\partial S}/{\partial N_i}$ in terms of the coefficients of \eqref{eq:FirstLawThermobis2}.  Using these identifications to compute, in particular, the crossed partial derivatives ${\partial^2 S}/{\partial T\partial V}$ and ${\partial^2 S}/{\partial V\partial T}$, which must be equal on account of the integrability condition of the exact differential form $dS$, we obtain after a straightforward calculation: $dp/dT=(\rho+p)/T$. Substituting this in $TdS=d[(\rho+p)V]-Vdp-\sum_i\mu_i dN_i$, which is just an appropriate rewriting of  Eq.\,\eqref{eq:FirstLawThermo}, we find:
\begin{equation}\label{eq:identif}
d S=\frac{1}{T} d\left[(\rho+p)V\right]-(\rho+p)\frac{V}{T^2}dT-\sum_i\frac{\mu_i}{T}\, dN_i=d\left[\frac{(\rho+p)V}{T}\right]-\sum_i\frac{\mu_i}{T}\, d(n_i V)\,,
\end{equation}
where $n_i=N_i/V$ is the number density of the ith species of particles.  Thus, in thermodynamical equilibrium, we find that the quantity
\begin{equation}\label{eq:Sconservat}
S=\frac{V}{T}\,(\rho+p-\sum_i\mu_i n_i)
\end{equation}
is conserved (up to an additive constant, which we can set to zero).  The obtained result is nothing but the total entropy contained in volume $V$ at equilibrium temperature $T$, and with chemical potentials adopting equilibrium values $\mu_i\, (i=1,2,...)$.  The corresponding entropy density is $s=S/V$, given by \eqref{eq:entropydensity}, as desired. Taking the coordinate volume equal to unity, as usually done\,\cite{KolbTurner:1990},  one may just work with the comoving volume $V=a^3$.

\mysection{Appendix B: Generalization to $H^{n+2}$-driven inflation}

Herein we study a more general class of RVM models, in which the structure of the dynamical cosmological term takes the form
\begin{equation}\label{eq:Lambdan}
\Lambda(H)=c_0+3\nu H^2+3\alpha \frac{H^{n+2}}{H_I^n}\,.
\end{equation}
The case $n=2$ is the one we have studied in the main text, see Eq.\,\eqref{eq:rhoLambdaunified}. Let us remark that, although odd values of $n=1,3,5,...$ would still trigger inflation through $H^3, H^5, H^7,...$ contributions of the vacuum energy density, the even values  $n=2,4,6,...$ are definitely preferred in practice since they involve an even number of derivatives of the scale factor through the higher order terms of the form  $H^4, H^6, H^8,...$ as being part of the vacuum energy density. Thus these terms are compatible with the general covariance of the effective action of QFT in curved spacetime (from which we should expect that the vacuum energy density should derive) and hence these terms are theoretically more favored for triggering inflation in generalized RVM frameworks, see e.g. \cite{JSPRev2013,Sola:2015rra}. The subsequent analytical calculations will be, though, for arbitrary $n$.

For general  $n\geq1$ (which will be assumed throughout) we can still ignore the last term $3\alpha \frac{H^{n+2}}{H_I^n}$ of Eq.\,\eqref{eq:Lambdan} whenever we consider the late universe, thus giving exactly the same phenomenology as in the case $n=2$ studied in the main text. Such low-energy phenomenology only depends on the first two terms on the \textit{r.h.s.} of Eq.\,\eqref{eq:Lambdan}.  The difference with the case $n=2$, therefore, appears only in the early universe, where the higher power term becomes dominant. For this reason we shall present here the results  for the early universe with arbitrary $n\geq1$ by  assuming that $c_0$ can be neglected. Our concern is whether different values of $n$ have a different impact on the entropy of the early universe.
\begin{figure}[t]
\begin{center}
\includegraphics[width=8cm,height=6cm]{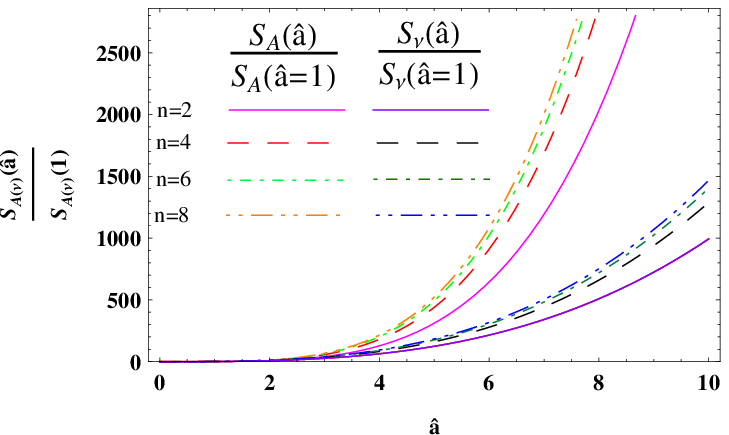}
\includegraphics[width=8cm,height=6.5cm]{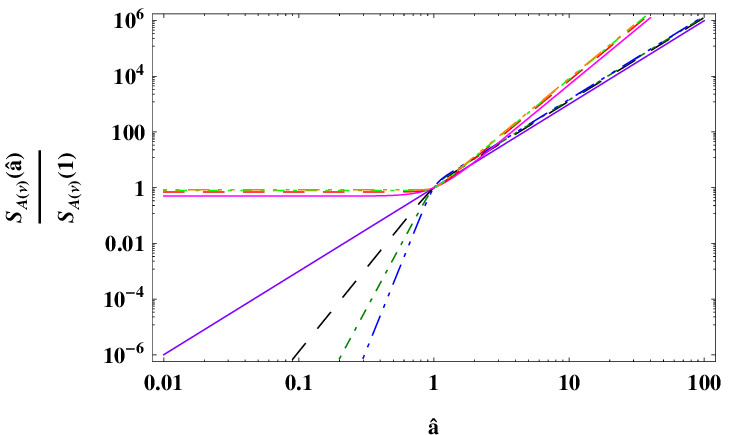}
\end{center}
\caption{\label{draft1} \scriptsize Evolution of the radiation entropy inside the horizon and the horizon entropy in the transit from the early universe to the radiation-dominated epoch. The left panel shows the radiation entropy and horizon entropy for different values of $n$ as a function of the (rescaled) scale factor $\hat a$, whereas the right one displays the same results in bilogarithmic scale in order to better appraise the region $\ha\to 0$ (similarly as in Fig\,4). The parameter $\nu$ is set to $0.001$ in all cases for different values of $n$:  $2$ (pink solid and purple solid), $4$ (red dashed and black dashed), $6$ (green dash-dotted and dark green dash-dotted), and $8$ (orange dash-dotted and blue dash-dotted). For each set of parameters, the first line stroke corresponds to the horizon entropy, and the second one to the radiation entropy.
}\label{final}
\end{figure}
By combining the field equations with the expression of $\Lambda(H)$, we can obtain the solution for the Hubble function. We do not repeat the detailed calculations here, which are more involved, so we just present the final results. For the Hubble function in the early universe, we find:
\begin{eqnarray}\label{hn1}
H(a)=\left(\frac{1-\nu}{\alpha}\right)^{1/n}\,\frac{H_I}{\left[1+D a^{\frac{3}{2} n (1-\nu) (1+\omega)}\right]^{1/n}}\,,
\end{eqnarray}
where $\omega$ will be henceforth set equal to $1/3$  since we are dealing with relativistic particles in the early universe. The solutions for the energy density of matter and vacuum read
\begin{eqnarray}\label{hn3}
\rho_r=\frac{3 H_I^2 (1-\nu)^{\frac{2}{n}+1}}{\kappa^2 \alpha ^{2/n} } \frac{Da^{2 n (1-\nu)}} {\left[1+D a^{2 n (1-\nu)}\right]^{\frac{n+2}{n}}}
\end{eqnarray}
and
\begin{eqnarray}\label{hn2}
\rho_{\Lambda}=\frac{3 H_I^2  (1-\nu)^{2/n}}{\kappa^2 \alpha ^{2/n}} \frac{1+\nu D a^{2 n (1-\nu)}}{\left[1+D a^{2 n (1-\nu}\right]^{\frac{n+2}{n}}}\,.
\end{eqnarray}
The above three formulas are the generalizations of equations \eqref{15}, \eqref{16.1} and \eqref{16.2}, respectively.
For convenience, we now follow once more the kind of notation adopted in Sec.~\ref{RVMearlyUniverse} and extend it for arbitrary $n$. We first find the equality point between vacuum energy density and relativistic matter, $a_{\rm eq}$, which satisfies
$\rho_r(a_{\rm eq})=\rho_{\Lambda}(a_{\rm eq})$. Thereafter we define the related point $\astar$ as follows:
\begin{eqnarray}\label{aeq00}
D=\frac{1}{1-2\nu}\,a_{\rm eq}^{-2n(1-\nu)}\equiv\astar^{-2n(1-\nu)}\,.
\end{eqnarray}
Next we introduce rescaled variables $\ha,  \tHI, \trI,  \tTI$ through the following relations, in a manner similar to the definitions given in Sec.~\ref{RVMearlyUniverse}:
\begin{eqnarray}\label{ha2}
\ha=\frac{a}{\astar}\,,\,\,\,\,\,\,\,\,\,\,\,\,\tHI=\left(\frac{1-\nu}{\alpha}\right)^{1/n}\,H_I\,\,\,\,\,\,\, \text{and} \,\,\,\,\,\,
\trI=\frac{3}{\kappa^2}\,\tHI^2 =\frac{\pi^2}{30}\,g_\ast\,\tTI^4\,.
\end{eqnarray}
 We are now ready to reexpress the above formulas in a much more compact way with the help of the those rescaled quantities.
First, we can rewrite Eq.~(\ref{hn1}) as
\begin{eqnarray}\label{hn11}
H(\ha)=\frac{\tHI}{\left[1+\ha^{2n(1-\nu)}\right]^{1/n}}.
\end{eqnarray}
Similarly for the matter and vacuum energy densities:
\begin{eqnarray}\label{eq:rhorfinaln}
\rho_r(\ha)&=&\trI (1-\nu)\,\frac{\ha^{2n(1-\nu)}}{\left[1+  \ha^{2n(1-\nu)}\right]^{\frac{n+2}{n}}}
\end{eqnarray}
and
\begin{eqnarray}\label{eq:rhoLfinaln}
\rho_\Lambda(\ha)&=&\trI\, \frac{1+\nu \ha^{2n(1-\nu)}}{\left[1+  \ha^{2n(1-\nu)}\right]^{\frac{n+2}{n}}}\,.
\end{eqnarray}

 We are ready to provide the general entropy formulas for arbitrary $n$.  First of all we compute the expression of the radiation temperature for arbitrary $n$:
\begin{eqnarray}\label{eq:Trn}
T_r =\tTI\,(1-\nu)^{1/4} \frac{\ha^{\frac{n}{2}(1-\nu)}}{\left[1+  \ha^{2n(1-\nu)}\right]^{\frac14(1+\frac{2}{n})}}\,.
\end{eqnarray}
We can easily check that for $\ha\gg1$ we find $T_r a^{1-\nu}={\rm const}.$, independent of $n$. Therefore, we recover the same law \eqref{eq:Tanuconst} for all $n$.
The evolution of the radiation entropy in the comoving volume is now easily found from \eqref{eq:Trn}, with the following result:
\begin{eqnarray}\label{eq:SrRVMn}
S_r(\hat{a})=\frac{2\pi^2}{45}\,g_{*}\,\tTI^3\,\astar^3\,f_n(\ha)\,.
\end{eqnarray}
Here we have defined the function
\begin{equation}\label{eq:fhan}
  f_n(\ha)=(1-\nu)^{3/4}\,\frac{\ha^{\frac{3n}{2}(1-\nu+\frac{2}{n})}}{\left[1+  \ha^{2n(1-\nu)}\right]^{\frac34(1+\frac{2}{n})}}\,,
\end{equation}
which generalizes that of Eq.\,\eqref{eq:fha} for arbitrary $n$. As expected,  for $n=2$ we recover the results of Sec.\,\ref{ProductionEarlyUniverse}.  After straightforward calculations it is not difficult to show the  following:  i) $S'(\ha)>0$ for all $\ha$; and ii)  for any value of $n$ there exists an inflexion point $\ha_i$  near $\ha=1$  such that  $S''(\ha)<0$ for $\ha>\ha_i$, i.e. a situation very similar to the behavior shown in Figs. 1 and 2 for the case $n=2$.   For arbitrary $n$, the rising of the entropy during the inflationary epoch (i.e. for  $\ha<1$) goes approximately (neglecting $\nu$ at this point)  as $S(\ha)\sim\ha^{3(1+n/2)}  $ and it can be extremely fast. For $n=2$ we recover $S(\ha)\sim \ha^6$, but e.g. for $n=4$ we have $S(\ha)\sim \ha^9$.  Deep into the radiation epoch  ($\ha \gg1$) it is easy to check that the raise of entropy once more  levels off and saturates to an asymptotic value, which is independent of  $n$ and therefore reads formally as in the $n=2$ case:
\begin{equation}\label{eq:SrSaturation2}
  S_r(\ha\gg1)\simeq \frac{2\pi^2}{45}\,g_{*}\,\tTI^3\astar^3(1-\nu)^{3/4}\ha^{3\nu}
  \equiv S_{r0} \ha^{3\nu}\,.
  \end{equation}
  The conclusions are therefore the same as for $n=2$, namely  for $0<\nu\ll1$ the behavior of the total entropy once more satisfies $S'_r(\ha)>0$ and $S''_r(\ha)<0$ for all $\ha\gg1$, in accordance with the Second Law.

Let us now focus on the GSL for arbitrary $n$  by considering the evolution of the total entropy inside and upon the (apparent) horizon, i.e. along the lines described in Sec.\,\ref{generalizedthermodynamiclaw} but considering arbitrary $n$. The total entropy is found to be
\begin{eqnarray}\label{hn4}
S_{total}(\ha)&=&\SV(\ha)+\SA(\ha)=\frac{\pi  M^2_P  \left[ 1+\hat{a}^{2n(1-\nu)} \right]^{2/n}}{\tHI^2}\nonumber\\
&+&\frac{8\pi^3}{135}\,g_{\ast}
\left(\frac{\tTI}{\tHI}\right)^3(1-\nu)^{3/4}\,
\ha^{\frac{3}{2} n (1-\nu)} \left[1+\ha^{2 n (1-\nu)}\right]^{\frac{3 (2-n)}{4 n}}\,,
\end{eqnarray}
which reduces to (\ref{24}) when $n=2$. The corresponding expression for $S_{total}'(\ha)$ reads as follows:
\begin{eqnarray}\label{hn5}
S_{total}'(\ha)&=&\frac{4\pi (1-\nu)}{45\tHI^3}\,\ha^{\frac{7}{2}n(1-\nu)-1}
[1+\ha^{2n(1-\nu)}]^{\frac{3}{2n}-\frac{7}{4}}\Big[45\tHI\, M_P^2\,\ha^{-\frac{3}{2}n(1-\nu)}
[1+\ha^{2n(1-\nu)}]^{\frac{3}{4}+\frac{1}{2n}}\nonumber\\
&+&\pi^2\,g_{\ast}\,\tTI^3
(1-\nu)^{3/4}(2+n\,\ha^{-2n(1-\nu)})\Big]\,.
\end{eqnarray}
For $n=2$ this expression boils down to \eqref{25}, as it should.
We find once more that $S_{total}'(\ha)>0$ for general $n$ on account of $0<\nu\ll1$, which is the same conclusion as in the case $n=2$.

As for $S_{total}''(\ha)$ the calculation is lengthier and we divide it into the two parts $\SV''(\ha)$ and $\SA''(\ha)$. The final results can be expressed as
\begin{eqnarray}\label{129}
\SV''(\ha)=&&\frac{2 \pi^3\,g_{\ast}\,\tTI^3}{45 \tHI^3} (1-\nu)^{7/4}\ha^{\frac{11}{2}n(1-\nu)-2} [1+\ha^{2 n(1-\nu)}]^{\frac{3}{2 n}-\frac{11}{4}}\nonumber\\ &&\Big[8-12\nu-(2n-3n^2+3n^2\nu)\ha^{-4n(1-\nu)}\nonumber\\
&&-[4+2n(2n-9+10\nu-2n\nu)]\ha^{-2n(1-\nu)}\Big],
\end{eqnarray}
and
\begin{eqnarray}
\SA''(\ha)=\frac{4 \pi  M_P^2 (1-\nu)\left[1+\ha^{2n(1-\nu)}\right]^{2/n}
\left[3-4\nu-(1+2n\nu-2n) \ha^{-2n(1-\nu)}\right]}
{\tHI^2\,\ha^2 \left[1+\ha^{-2 n (1-\nu)}\right]^2}.
\end{eqnarray}
One can check that the sum of these more complicated formulas reduces to Eq.\,\eqref{252} for $n=2$.
Furthermore, it is easy to check that when $\ha$ tends to 0, $\SV''(\ha)$ and $\SA''(\ha)$ are both positive in the considered range of values. Therefore, $S_{total}''(\ha)$ is always positive in the very early universe, which is consistent with the conclusion obtained in the case  $n=2$. In other words, the entropy of the early universe has an ``explosive'' growth regardless of the value of $n$. As we have explained in the main text, this result is acceptable since the behavior of the entropy is rectified at late times. Our main concern is whether the total entropy $S_{total}(\ha)$ can eventually reach equilibrium, i.e.  $S_{total}''(\ha\gg1)<0$. From Eq.~(\ref{129}), we  find that $\SV''(\ha\gg1)$ is positive since $0<\nu\ll1$. Similarly, we can prove that  $\SA''(\ha)>0$ holds true for $\ha>0$. Moreover, one can check explicitly that the  asymptotic behaviors of the above expressions do not depend on $n$:
\begin{eqnarray}
\SV''(\ha\gg1)\sim \ha^{1-3\nu}\,\,\,\,\,\,\,\,\,{\rm and}\,\,\,\,\,\,
\SA''(\ha\gg1)\sim \ha^{2-4\nu}\,.
\end{eqnarray}
Thus, they turn out to be equal to the respective asymptotic behaviors of the two terms of Eq.\, (\ref{252}) corresponding to $n=2$. As a matter of fact, it could be expected that the asymptotic behaviors of $\SV''(\ha)$ and $\SA''(\ha)$ should be independent of the value of $n$, because with the growth of the scale factor, the term $3\alpha \frac{H^{n+2}}{H_I^n}$ in $\Lambda(H)$ will be eventually negligible.

In Fig.~\ref{final} we draw plots displaying the exact numerical evolution of the radiation entropy inside the horizon, together with the horizon entropy, for a few values of $n$, which we take all even for the reasons explained in the beginning. From that figure we can see that the larger the value of $n$, the faster is the corresponding entropy growth.

Finally, the following remark is in order. The evolutions of all the cosmological functions, such as the scale factor, vacuum energy density and radiation entropy, are continuous functions of the cosmic time. If we take into account that for general $n$ the evolution of radiation entropy in the late universe is consistent with the case $n=2$, we are also guaranteed that the thermodynamic requirements of the Second Law and of the LTE ($S'(a)>0$ and $S''(a)<0$) must also be fulfilled for general $n$, much in the same way as we have explicitly shown for the particular case $n=2$. Once more we see that the argument hinges on the necessary presence of the nonvanishing (positive) term $c_0$ in the structure of the vacuum energy density Eq.\,\eqref{eq:Lambdan}. In this way we have formally proven that all of the generalized RVM models (\ref{eq:Lambdan}) follow the same thermodynamical pattern, irrespective of the value of $n\geq1$, and therefore they all fulfill the GSL.  In all these cases the existence of a positive cosmological constant  is instrumental to warrant a safe path of the universe towards stable thermodynamical equilibrium.
\newpage

\end{document}